\documentclass[nologo,11pt,a4paper]{ETHpaper}
\usepackage{graphicx, amsmath, amssymb,color,wasysym,amsthm}
\usepackage[square,numbers,sort&compress]{natbib}
\usepackage{longtable}
\usepackage{booktabs}
\usepackage{mathtools}
\usepackage{cancel}

\usepackage{cleveref}
\usepackage{caption}
\usepackage{subcaption}
\usepackage{multicol}

\begin{document}

\newcommand{\mean}[1]{\left\langle #1 \right\rangle} \newcommand{\abs}[1]{\left| #1 \right|}

\title{Improving the robustness of online social networks: A simulation approach of network interventions}

\titlealternative{Improving the robustness of online social networks: A simulation approach of network interventions}

\author{Giona Casiraghi, Frank Schweitzer
  }

\authoralternative{G. Casiraghi, F. Schweitzer} 
\address{Chair of Systems Design, ETH Zurich, Weinbergstrasse 58, 8092 Zurich, Switzerland}

\reference{(Submitted)}

\www{\url{http://www.sg.ethz.ch}}

\makeframing
\maketitle

\begin{abstract}
 Online social networks (OSN) are prime examples of socio-technical systems in which individuals interact via a technical platform.
 OSN are very volatile because users enter and exit and frequently change their interactions.
 This makes the robustness of such systems difficult to measure and to control.
 To quantify robustness, we propose a coreness value obtained from the directed interaction network.
 We study the emergence of large drop-out cascades of users leaving the OSN by means of an agent-based model.
 For agents, we define a utility function that depends on their relative reputation and their costs for interactions.
 The decision of agents to leave the OSN depends on this utility.
 Our aim is to prevent drop-out cascades by influencing specific agents with low utility.
 We identify strategies to control agents in the core and the periphery
  of the OSN such that drop-out cascades are significantly reduced, and the robustness of the OSN is increased.

  \end{abstract}
\date{\today}

\section{Introduction}
\label{sec:Introduction}

Self-organization describes a \emph{collective dynamics} resulting from the \emph{local   interactions} of a vast number of system elements \citep{fs-ed-97}, denoted in the following as \emph{agents}.
 The macroscopic properties that \emph{emerge} on the system level are often desired, for example, coherent motion in swarms or functionality in gene regulatory networks.
 But as often these self-organized systemic properties are \emph{not} desired, for example, traffic jams or mass panics in social systems.
 Hence, while self-organization can be a very useful dynamics, we need to find ways of controlling it such that systemic malfunction can be excluded, or at least mitigated.
 This refers to the bigger picture of \emph{systems design} \citep{Schweitzer:2019vp}: how can we influence systems in a way that optimal states can be achieved and inefficient or undesired states can be avoided?

In general, self-organizing processes can be controlled, or designed, in different ways.
 On the \emph{macroscopic} or systemic level, \emph{global control parameters}, like boundary conditions, can be adjusted such that phase transitions or regime shifts become impossible.
This can be done more easily for physical or chemical systems, where temperature, pressure, chemical concentration, etc. can be fixed.
 On the \emph{microscopic} or agent level, we have two ways of controlling systems: (i) by influencing agents directly, (ii) by controlling their interactions.

Referring to socio-economic systems, we could, for example, incentivize agents to prefer certain options, this way impacting their utility function.
 This requires to have access to agents, which is not always guaranteed.
 For instance, it is difficult to access prominent agents or to influence large multi-national companies.
 Controlling agents' interactions, on the other hand, basically means to restrict (or to enhance) their communication, i.e., their access to information and dissemination.
 Restrictions can be implemented both globally and locally.

In this paper, we address one particular instance of social systems, namely \emph{online social networks} (OSN).
 Prominent examples for such networks are \texttt{facebook}, \texttt{reddit}, or \texttt{Twitter}.
 OSN are instances of a complex system comprising a large number of interacting agents which represent users of such networks.
OSN are, in fact, \emph{socio-technical systems} because they combine elements of a social system, i.e., users communicating, with elements of a technical system, i.e., platforms, protocols, GUI (graphical user interfaces), etc.
 The technical component is important because it allows to \emph{control} the access to users, as well as their communication.
 The term \emph{control} refers to the fact that access and interactions are \emph{monitored}, but also \emph{influenced} in different ways.

In reality, it becomes very difficult to control OSN because of their large \emph{volatility}, which has two causes.
The first one is the \emph{entry and exit} dynamics, which impacts the number of \emph{agents}:  Users enter or leave the OSN at a high frequency.
The second one is the \emph{connectivity}, which impacts the number of \emph{interactions}: Users easily connect to and disconnect from other users or interact with lower or higher frequency.
They have ample ways of interacting; thus, it becomes very difficult to shield them from certain information.

Because of this volatility,  in an OSN \emph{interactions} cannot be fully controlled.
But we can certainly influence users via their \emph{utility function}.
 Users join an OSN for a certain purpose, namely to socialize and to exchange information.
 Hence, their benefits are a function of the number of other users they interact with.
 Their costs, on the other hand, result from the effort of maintaining their profile, learning about the features of the graphical user interface, etc.
 The utility, i.e., the difference between benefits and costs, can then be increased by either increasing the benefits, e.g., by increasing their number of friends, or by decreasing their costs, e.g., by automatizing profile updates, or by a combination of both.

OSN are a paradigm for the emergence of collective dynamics and are much studied because of this.
For example, the emergence of trends, fashions, social norms, or opinions occurs as a self-organized process that can sometimes be initiated but hardly be controlled.
 A worrying trend emerges if users decide to \emph{leave} the social network.
 If their decision causes other users to leave as well, because they lost their friends, this can quickly result in large drop-out cascades and in the total collapse of the OSN~\cite{Kairam2012}.
 This happened, for example, to \emph{friendster}, an OSN with about 117 million users in 2011.
 As studied in detail \citep{Garcia2013}, less integrated users left \texttt{friendster}, this way, making it less attractive to the remaining users to further stay on the platform.

To model such a self-organized dynamics by means of an agent-based model requires us to solve a number of methodological issues.
 On the \emph{agent} level, we need to model individual decisions of agents based on their perceived utility, which is to be defined.
 On the \emph{system} level, we need to quantify how the drop-out of individual agents impact other agents and the whole system, in the end~\cite{Jain1998,Jain2002}.
In a volatile system, agents come and go at a large rate, without threatening the stability of the system every time.
 Hence, we need to define a macroscopic measure that allows quantifying whether the system is still robust.

Once these methodological issues are solved, we can turn to the more interesting question of \emph{systems design}.
 This means that, by using our agent-based model, we explore possibilities to influence the system such that it becomes more robust.
 Our focus will be on the \emph{microscopic} level, i.e., 
influencing \emph{agents} rather than whole systems.
 This is sometimes referred to as \emph{mechanism design}.
 But, different from designing communication, i.e., influencing \emph{interactions}, here we influence agents via their utility functions.
 This leads to another methodological problem, namely how to identify those agents that are worth to be influenced, i.e., are most promising for reaching a desired system state.

This problem is for networks addressed in the so-called \emph{controllability theory} \citep{Liu2011}, which is very much related to control theory in engineering.
 It allows to quantify how much of a network is controlled by a given agent, which then can be used to \emph{rank} agents with respect to their \emph{control capacity} \citep{zhang2019control}.
 To apply this formal framework, however, requires to have a static network, i.e., the interaction \emph{topology} should \emph{not} change on the same time scale as the interaction.
 So, this framework does \emph{not} allow us to study drop-out cascades in which the network topology changes at every time step.
 Because of this, in our paper, we have to rely on a \emph{computational approach}.
 i.e., we use our agent-based model to simulate the decision of agents to leave the network and its impact on the remaining network, while monitoring the overall robustness of the system by means of a macroscopic measure.

With these considerations, we have already specified the structure of this paper.
 In Sect.~\ref{sec:robustn-soci-netw}, we model the decisions of agents and quantify the robustness of the network.
 In Sect.~\ref{sec:reputation-dynamics}, we introduce a reputation dynamics that runs on the network, to determine the benefits of the agents.
In Sect.~\ref{sec:network-dynamics} we highlight the dynamics of the OSN without any interventions, to demonstrate its breakdown.
 In Sect. \ref{sec:improving-robustness}, eventually, we use our model to explore different agent-based strategies of improving the robustness of the network.

\section{Robustness of the social network}
\label{sec:robustn-soci-netw}

\subsection{Agents and interaction networks}
\label{sec:agents-inter-netw}

\paragraph{Networks. \ }

For our agent-based model of the OSN we use the specific representation of a \emph{complex network}.
The term \emph{complex} refers to the fact that we have a large number of interacting agents such that new system properties can \emph{emerge} as the result of these collective interactions.
The term \emph{network} means that agents are represented by \emph{nodes}, and their interactions by \emph{links} of the network.  This implies that all interactions are decomposed into dyadic interactions between any two agents.

Using a mathematical language, networks are denoted as
\emph{graphs}, nodes as \emph{vertices} and links as \emph{edges}.
We can then formally define a \emph{graph} object $\mathcal G$ as an ordered pair $\mathcal G=\mathcal G(V,E)$, where $V$ is the set of vertices of the graph, and $E$ is the set of edges. 
Vertex $i\in V$ and $j\in V$ are connected if and only if $ij\in E$.
The graph is not static but changes on a time scale $T$, i.e., $\mathcal G(T)$.  We call $T$ the \emph{network time} because agents can enter or exit the OSN, this way changing both the number of vertices and edges.

Agents are characterized by an binary state variable $s_{i}(T)\in \{0,1\}$, where
$s_{i}(T)=1$ means that agent $i$ at time $T$ decides to \emph{stay} in the OSN, whereas
$s_{i}(T)=0$ means that it decides to \emph{leave} the OSN.  This decision is governed by a utility
function $U_{i}(T)$:
\begin{equation}
  \label{eq:choice}
  s_{i}(T):=\Theta[U_{i}(T)] \;; \quad U_{i}(T)= B_{i}(T)-C_{i}(T)
\end{equation}
The Heaviside function $\Theta(x)$ returns $1$ if $x\geq0$ and $0$ otherwise.  $B_{i}(T)$ and
$C_{i}(T)$ are the benefits and the costs of agent $i$ at time $T$.  Only if the benefits exceed the
costs, agent $i$ will stay in the OSN, otherwise it leaves.  The two functions need to be further
specified, which is done in Sect. \ref{sec:reputation-dynamics}.

\paragraph{Interactions. \ }

We want to model an OSN; therefore, we consider \emph{directed interactions} between agents.  Taking
the example of \texttt{Twitter}, a directed interaction $i\to j$ means that agent $i$ is a follower
of agent $j$.  Obviously, the reverse does not need to apply but can be frequently observed.  Each
of these interactions is represented as a directed link in the network $\mathcal{G}$.  A formal
expression for the topology of a network with $N$ agents is the \emph{adjacency matrix}
$\mathcal A\in\mathbb N^{N\times N}$ in which the elements $a_{ij}$ are either $0$ or $1$.  This
allows to define the \emph{in-degree} $d^{+}_{i}$ and the \emph{out-degree} $d^{-}_{i}$ of an agent
$i\in V$ as the number of incoming or outgoing links of $i$.  We can also define the \emph{total
  degree} of agent $i$ as the sum of both in- and out-degree, $d_{i}=d^{+}_{i}+d_{i}^{-}$.

The degree distribution is an important macroscopic quantity to characterize a complex network.  It
is known that OSN have a rather broad degree distribution \citep{Garcia2013}, i.e., many agents are linked to only a few
other agents, while a few agents, called hubs, have very many incoming links from other agents.
Additionally, OSN often show a so-called \emph{core-periphery structure}~\citep{borgatti2000models}, in which well connected
agents form a core, whereas agents with only a few, or even no, connections form the periphery.
Identifying such structures helps to analyze the robustness of the network.  Precisely, we can
assume that the OSN is robust, despite an ongoing entry and exit of agents, if the core changes, but
continues to exist.  This implies that the volatile dynamics mostly affects the periphery.  If,
however, the drop-out of a few agents is amplified into a large drop-out cascade that affects even
the core of the OSN, then the robustness of the system is very low.  We need to come up with a
robustness measure that reflects such a situation appropriately.  This is developed in the next
section.

\subsection{Quantifying Robustness}
\label{sec:quant-robustn}

\paragraph{Coreness. \ }

We decided to use the \emph{coreness} $k_{i}$ of agents as our starting point because it reflects
from a topological perspective how well an agent is integrated into the network
\citep{Seidman1983a}.  A coreness value $k_{i}$ allows quantifying the impact on the network when
removing agent $i$.  Individual coreness values are obtained by means of a pruning procedure, which
is known as \emph{$k$-core decomposition}.  It assigns agents to different concentric \emph{shells} that reflect the integration of these agents in the network.  Specifically, the $k$-\emph{core} is
identified by subsequently pruning all agents with a degree $d_{i}$ less than $k$.  Pruning starts
with $k=1$ and stops when all the agents left have a degree greater or equal to $k_{\mathrm{max}}$.
The corresponding $k$-\emph{shell} then consists of all agents that are in a $k$-core but not in the
$(k+1)$-core, i.e.  agents assigned to a $k$-shell have coreness value $k_{i}=k$.

\begin{figure}
  \centering \includegraphics[width=.9\textwidth]{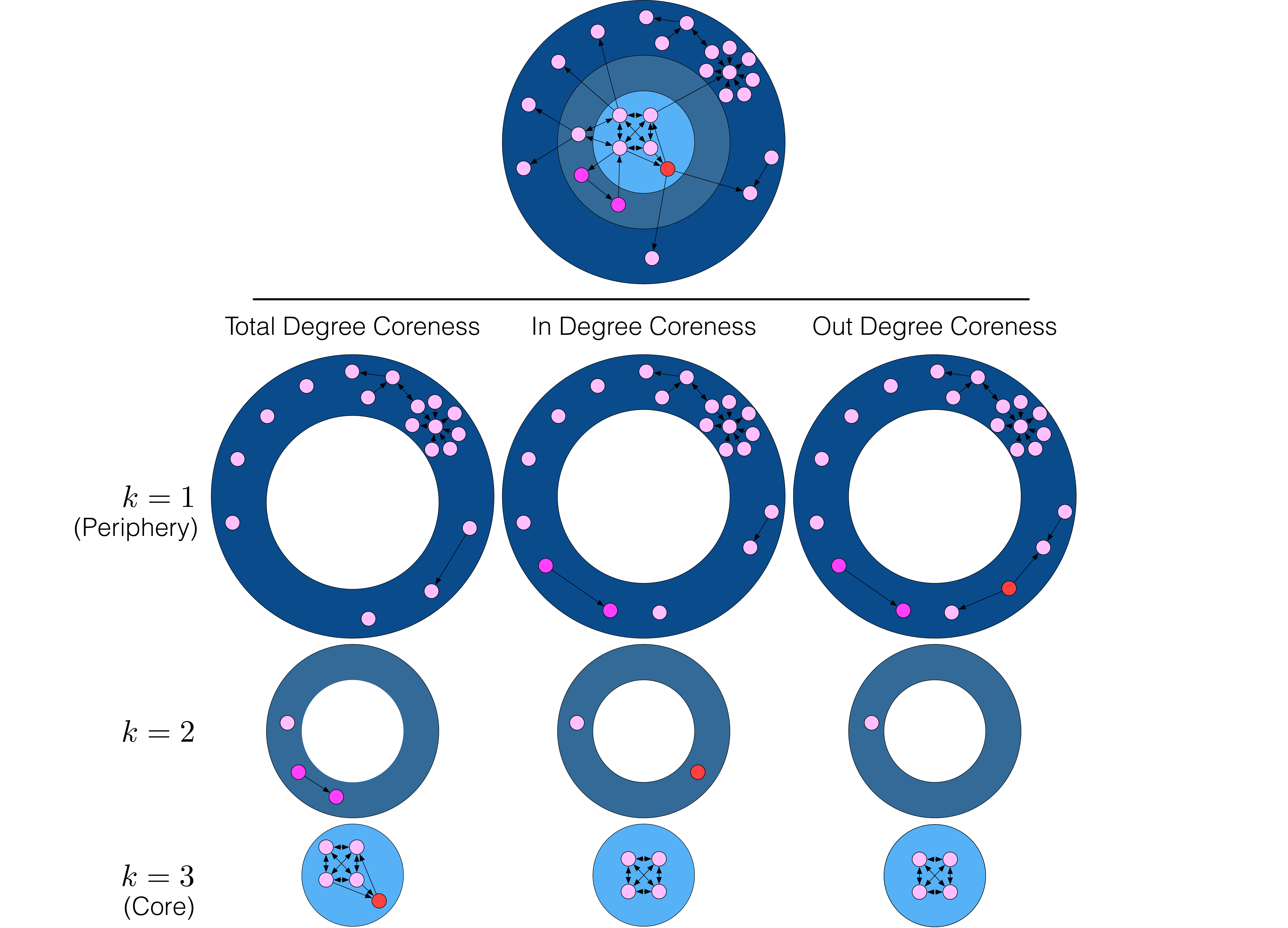}
  \caption{$k$-core decomposition of a network with 10 agents.}
  \label{fig:kcore}
\end{figure}

\Cref{fig:kcore} provides an illustration of the $k$-core decomposition applied to a network of 10 agents.  Agents with a coreness $k_{i}=1$ are located in the \emph{periphery} (dark blue), i.e., they are loosely
connected with the core.  Note that some of these agents have a relatively high
degree, in spite of their low coreness.  Agents with a coreness $k_{i}=2$ are closely connected to, but \emph{not yet} fully integrated into the core, belong to an intermediate shell (blue).  The 5 agents with coreness $k_{i}=k_{\mathrm{max}}=3$ are the most densely connected ones in this sample
network and belong to the innermost core (light blue).  This illustrates that the higher the coreness
$k_i$ of an agent $i$, the stronger the impact on the network when removing $i$ because this
potentially disconnects a large number of agents with lower coreness from the network.  Conversely,
removing agents with low coreness will have a weaker impact on the network because they belong to
outer shells, and removing them disconnects a smaller number of agents.

In this article, we want to quantify how much the drop-out of agents will impact the \emph{robustness} of the network.  As motivated above, robustness shall
be characterized by the \emph{average coreness} of the agents:
\begin{align}
  \label{eq:2}
  \mean{k}=\frac{1}{N}\sum_{k=1}^{k_{\mathrm{max}}}k\, n_{k}\;;\quad \sum_{k=1}^{k_{\mathrm{max}}}n_{k}=N
\end{align}
where $N=\abs{V}$ is the total number of (connected and disconnected) agents in the network and
$n_{k}$ is the number of agents with a coreness value $k_{i}=k$.  $\mean{k}$ will be high if either
most agents have a relatively high coreness, or few agents have a very high coreness.  In both
cases, the core of the network is less likely to be affected by cascades that started in the
periphery.  So, $\mean{k}$ summarizes the information we are interested in.  In this paper, we do
not focus on the \emph{heterogeneity} of coreness values, which could be described by the
\emph{variance} of the coreness distribution, or by \emph{coreness centralisation} \citep{Wasserman1994}.

\paragraph{In-degree and out-degree coreness. \ }
\label{sec:degree-out-degree}

The above definition of coreness is based on the total degree $d_{i}$ of agents, i.e., it is
appropriate for \emph{undirected} networks.  For the case of a \emph{directed} network discussed in
this paper, this may give wrong conclusions about the embeddedness of agents.  Therefore, we now
introduce two separate measures, in-degree coreness, $k_{i}^{+}$, and out-degree coreness,
$k_{i}^{-}$, which reflect the existence of directed links via the in- and out-degrees $d_{i}^{+}$,
$d_{i}^{-}$.

The results for the different metrics and the differences between them are illustrated in the sample
network of 10 agents in Figure \ref{fig:kcore}.  This network is characterized by 3 $k$-shells, but
it is important to note that the three different coreness metrics possibly assign the same agents to
very different $k$-shells.  Take the example of the pair of purple agents that, according to
total-degree coreness, are assigned to the shell $k=2$.  If we account for directionality of the
links, they are now assigned to $k=1$, i.e., to the \emph{periphery}.  Moreover, the red agent that,
according to the total degree coreness, belongs to the core, $k_{\mathrm{max}}=3$, is now assigned
to the shell $k=2$ if \emph{in-degree coreness} is taken into account, and to $k=1$, i.e. to the
periphery, if \emph{out-degree coreness} is instead considered.

This example makes clear that it very much depends on the \emph{application} whether coreness should
be calculated based on directed or undirected links, and whether in- or out-degrees should be
considered.  In the following we will use \emph{in-degree coreness}, $k_{i}^{+}$, to compute the
average coreness $\mean{k}$, Eqn.~\eqref{eq:2}, i.e. $n_{k}$ is the number of agents with in-degree
coreness $k_{i}^{+}=k$.  The reason for this choice comes from the benefits of agents defined in
\cref{eq:choice} and is discussed in the following section.

\section{Dynamics \emph{on} the social network}
\label{sec:reputation-dynamics}

\subsection{User benefits and costs}
\label{sec:user-benefits}

To enable a network dynamics on the time scale $T$, where agents can \emph{leave} the network
according to Eqn. \eqref{eq:choice}, we need to further specify their benefits, $B_{i}(T)$, and
costs, $C_{i}(T)$.  This leads to the question of why, in the real world, users join or leave an OSN.
There are certainly different reasons, such as information exchange, maintaining friendship links, or
receiving \emph{attention}.  From this, we can deduce that benefits should \emph{increase} with the
\emph{in-degree} $d^{+}_{i}$ of an agent in a monotonous, but likely non-linear manner.  For
instance, on \texttt{Twitter} attention increases with the \emph{number} of followers.  More
important, however, is not just the number, but also the \emph{importance} of the followers.  The
attention for a user $i$ can considerably increase if it has a number of important users $j$ following.  This amplifies the attention because, in an OSN, other users following the important user
$j$ this way also receive information from $i$.

To capture such effects in our agent-based model, we assign to each agent a second state variable,
\emph{reputation} $R_{i}$, which is continuous and positive.  In real-world OSN, user reputation
plays an important role and can be proxied by different measures, such as number of \texttt{likes}
in \texttt{Facebook} positive votes in \texttt{Amazon} and \texttt{Dooyoo}, or retweets on
\texttt{Twitter}.  Other proxies take the activity of users into account, for example, the RG score
from \texttt{Researchgate}, or the Karma points from \texttt{Reddit}.  All of these measures have the
drawback that they are (i) specific to the OSN, (ii) depend on the subjective judgment of other
users.

Therefore we resort to so-called \emph{feedback centrality} measures, prominently known from the
early versions of the \texttt{PageRank} algorithm, in which the importance (centrality) of a node in
a network entirely depends on the importance of the nodes linked to it.  This leads to a set of
equations for the importance of all nodes that has to be solved in a self-consistent way.  While
this is a crucial element to define our reputation measure, it is not enough to explain reputation.
We also need to consider that reputation fades out over time if it is not continuously
\emph{maintained}.  Usually, the reputation of an agent can be maintained in different ways, (i) by
the own effort of the agent and (ii) by means of direct interactions with others.  Such
considerations have been formalized in other reputation models \citep{Perony2019}.  Here, we only consider
the increase of reputation coming from other agents, to simplify the formalization.

In the following section, we will specify our dynamics for the reputation of an agent, which leads
to a stationary value of $R_{i}(T)$.  Given that we have calculated this value, we posit that the benefit of an
agent from being in the OSN comes from its reputation as a good proxy of the attention that this
agent receives from others.  The absolute value of $R_{i}$ will also depend on the network size and
the density of links.  What matters in an OSN is not the \emph{absolute} value, but the reputation of users \emph{relative} to that of
others.  Therefore we define the benefit $B_{i}$ for each agent $i\in V(\mathcal G)$ as the absolute
reputation rescaled by the largest reputation value $R_{\text{max}}(T)$ at the given time $T$.
\begin{equation}
  \label{eq:benefit}
  B_{i}(T):=b\frac{R_{i}(T)}{R_{\text{max}}(T)}=b \frac{R_{i}(T)}{\max_{j\in V(\mathcal G)}{R_{j}(T)}}
\end{equation}
The constant $b$ allows to weight the benefits from the reputation against the costs.

To specify the costs $C_{i}(T)$, in our model, we consider two contributions.  First, there are fixed
costs per time unit, $c_{0}$, that do not depend on the activity of the agents.  They capture, in a
real OSN, the minimal effort made by users to be present in the OSN, i.e., to learn about the GUI
and to maintain the profile.  The second contribution comes from the costly interaction with other
agents.  Because, for instance on \texttt{Twitter}, agent $i$ can only control whom to follow, these
costs should be proportional to the \emph{out-degree} $d^{-}_{i}$ of the agent, $c_{i}\,d^{-}_{i}$.
In a real OSN, the costs per interaction, $c_{i}$, are not the same for all users.  More prominent
users have, for example, much more time constraints because of other activities that compete for
their attention.  Therefore, it is reasonable to assume that $c_{i}$ is a non-linear function of the
user's reputation, $c_{i}(R_{i})=c_{1} R_{i}^{2}$.  The non-linearity induces a stronger saturation
effect for more prominent users in interacting with many other users.

As with the benefits, also the costs should not depend on the absolute reputation of the agent, but
on the relative one.  This leads to
\begin{equation}
  \label{eq:cost}
  C_{i}(T):=c_{0} + c_{1} d_{i}^{-}\left[\frac{R_{i}(T)}{R_{\text{max}}(T)}\right]^{2}.
  \end{equation}
Denoting the relative reputation at a given time $T$ as $r_{i}(T)={R_{i}(T)}/R_{\text{max}}(T)$, we
can eventually write down the utility function of agent $i$, Eqn.~\eqref{eq:choice}, as:
\begin{equation}
  \label{eq:1}
  U_{i}(T)= b\, r_{i}(T) - c_{0} - c_{1}d_{i}^{-}\, r_{i}^{2}(T) = -c_{0} +
  \left[b- c_{1}d_{i}^{-}\, r_{i}(T)\right]r_{i}(T).
\end{equation}

\subsection{Reputation dynamics}
\label{sec:reputation-dynamics-1}

After linking the utility function of agents to their reputation, we have to specify how to
calculate the latter.  In accordance with the above discussion, we use the following reputation
dynamics:
\begin{equation}
  \label{eq:singlenode}
  \frac{d R_{i}(t)}{d t}=-\gamma R_{i}(t) + \sum_{j\in V[\mathcal G(T)]}{a_{ji}R_{j}(t)}
\end{equation}
Here, $t$ denotes a \emph{time scale} much shorter than the time scale $T$ at which agents decide
whether to stay or to leave the OSN.  Hence, compared to the change of the \emph{network}, the
change of \emph{reputation} is \emph{fast} enough such that a stationary value $R_{i}(T)$ is
obtained at time $T$.

The first term in Eqn.~\eqref{eq:singlenode} expresses a continuous decay of reputation with a rate
$\gamma$, to reflect the fact that reputation fades out over time if it is not maintained.  The
second term captures the increase of reputation coming from other agents linked to agent $i$,
i.e., $a_{ji}=1$.  The summation is over all agents part of the OSN at time $T$.

\begin{figure}[htbp]
  
  \begin{subfigure}{.3\textwidth}
    \includegraphics[height=3cm]{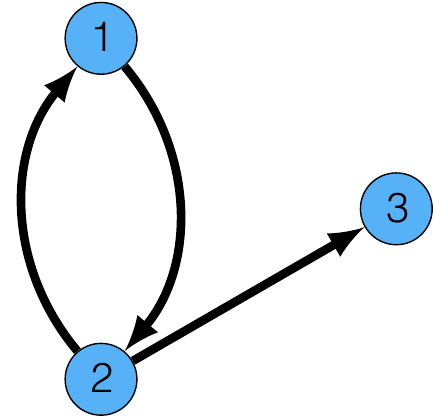}

    \centerline{\includegraphics[width=.99\textwidth,angle=0]{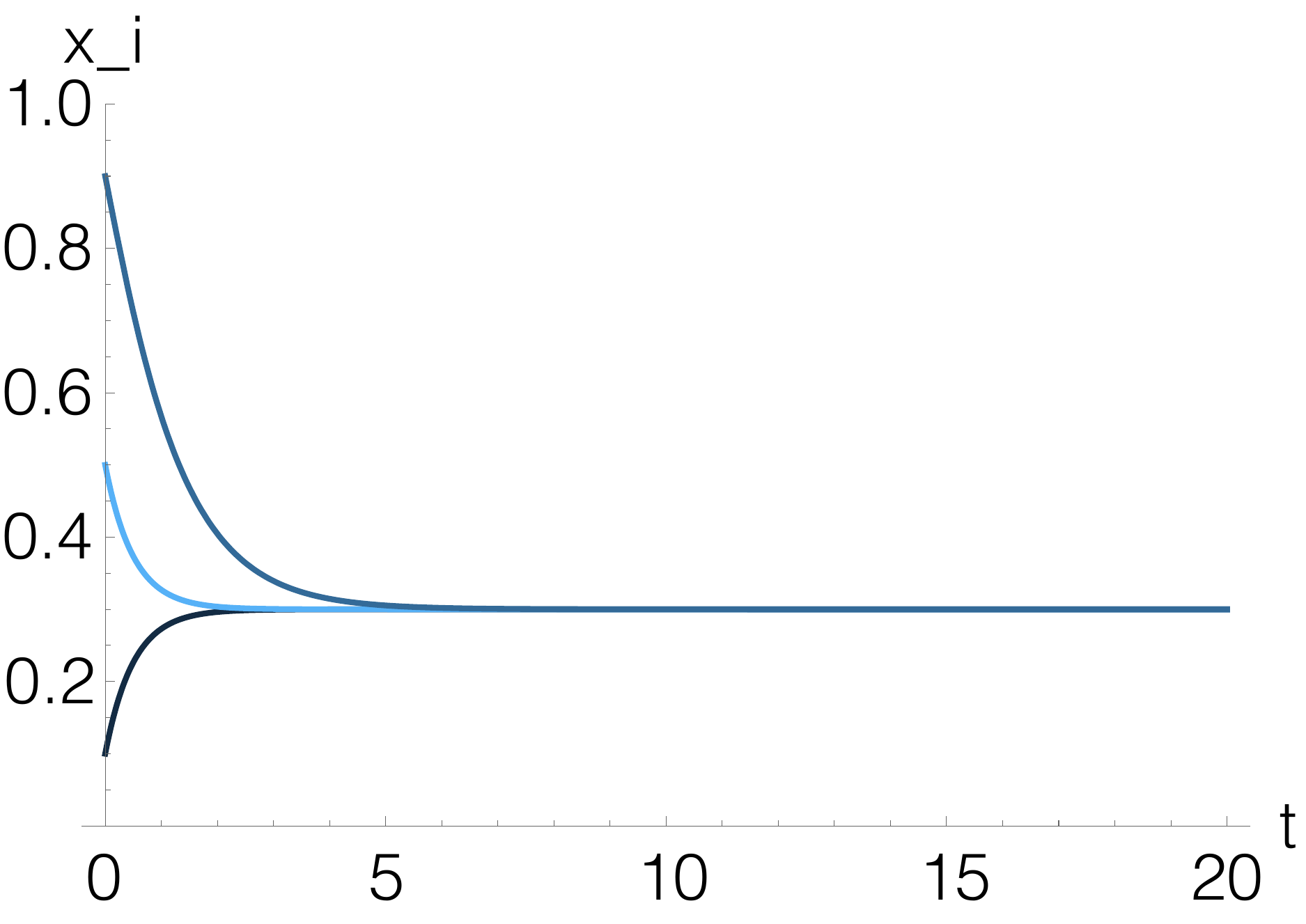}} \caption{}
  \end{subfigure}\hfill\vline\hfill
  \begin{subfigure}{.3\textwidth}
    \includegraphics[height=3cm]{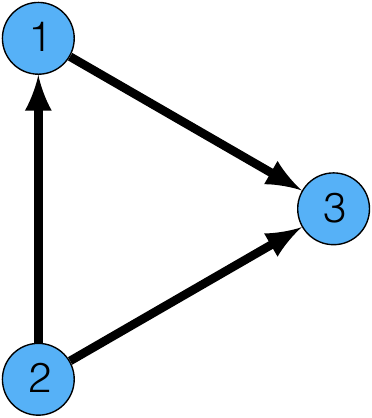}
    
    \centerline{\includegraphics[width=0.99\textwidth,angle=0]
        {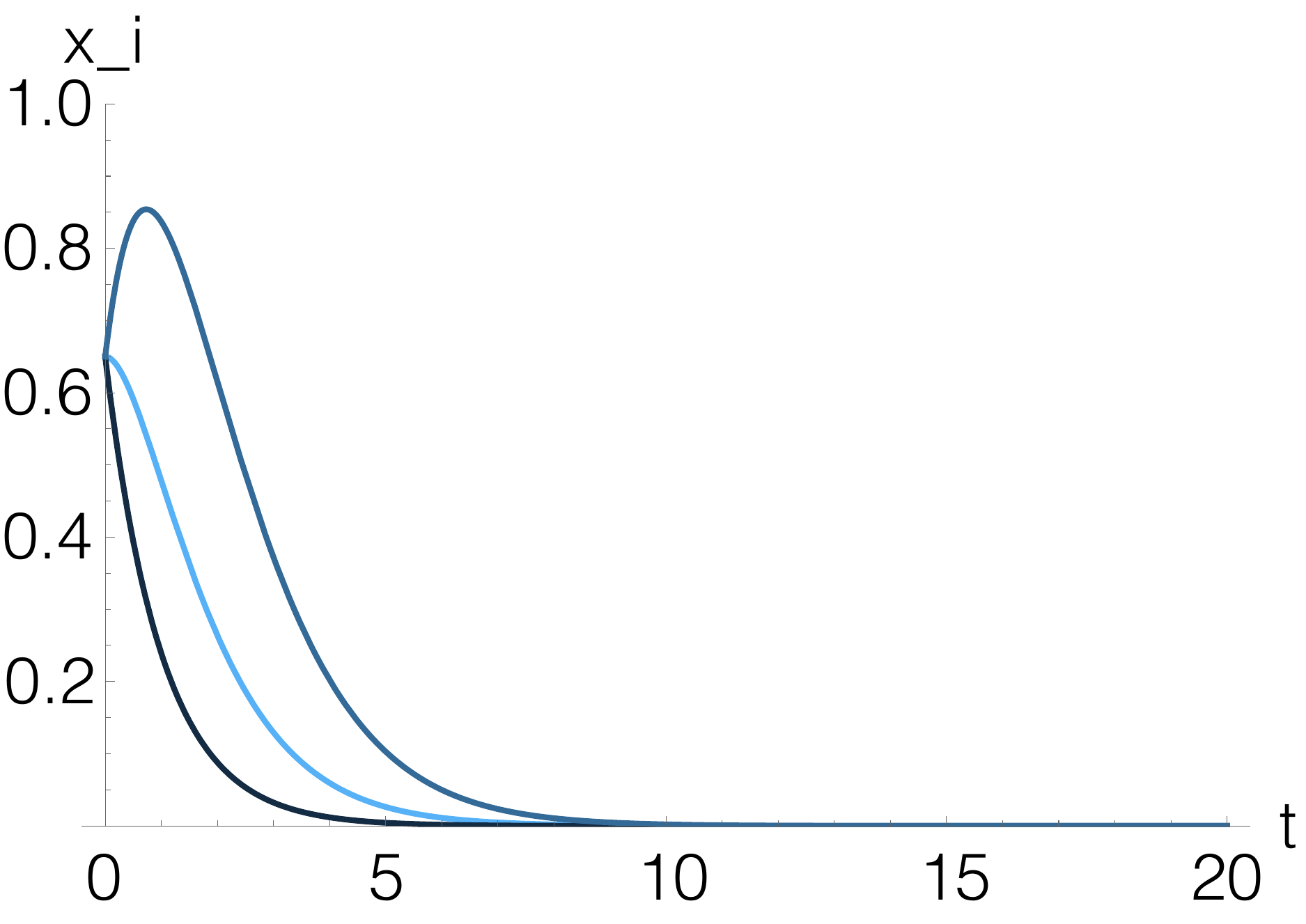}} \caption{}
\end{subfigure}\hfill\vline\hfill
\begin{subfigure}{.3\textwidth}
 \includegraphics[height=3cm]{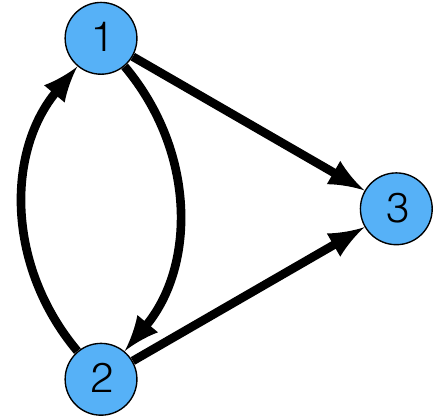}

    \centerline{\includegraphics[width=0.99\textwidth,angle=0]
        {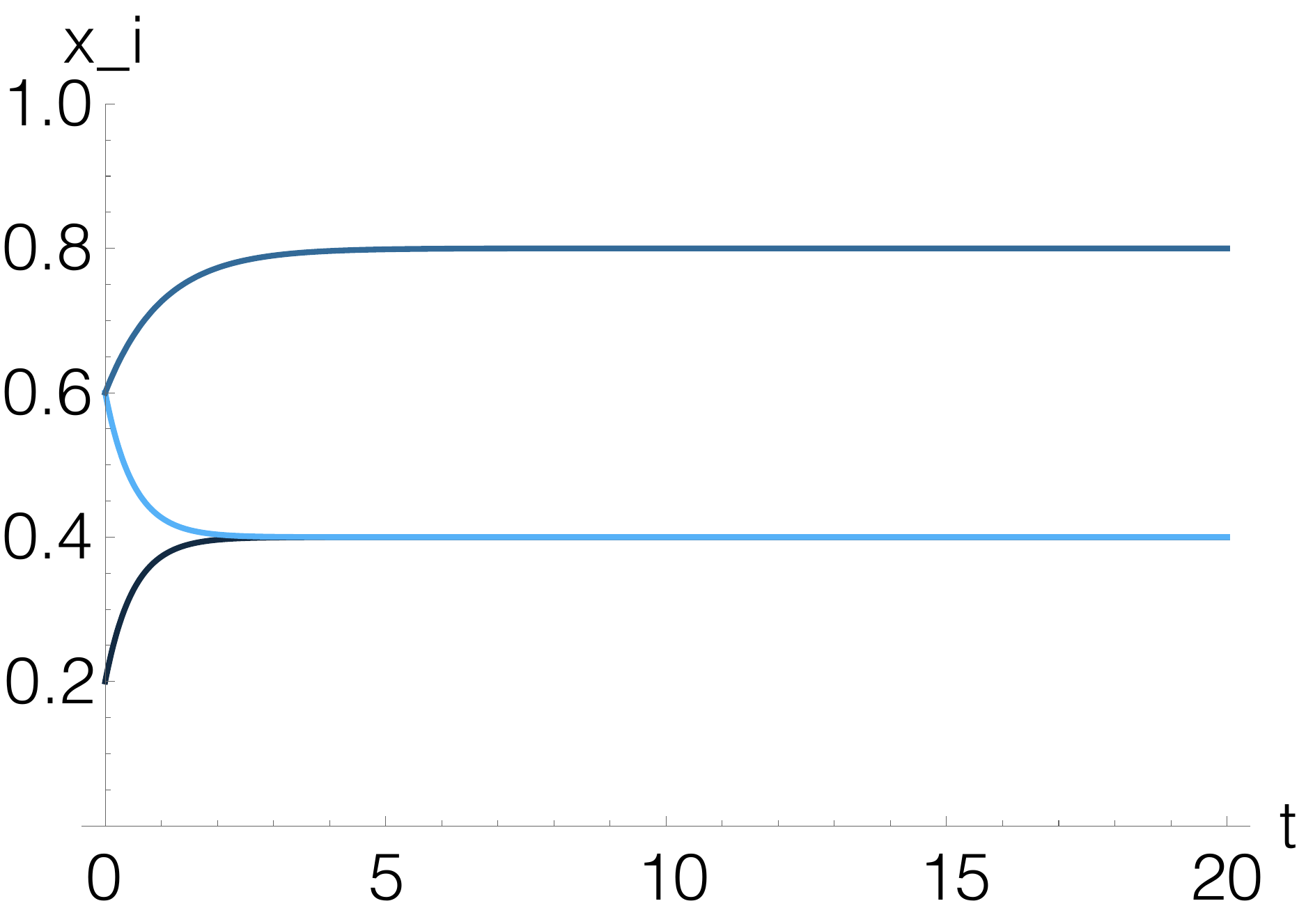}} \caption{}
\end{subfigure}
\caption{Impact of the adjacency matrix on the reputation $R_{i}(t)$ of three agents.
  Only if cycles exist and agents are connected to these cycles, a non-trivial stationary reputation can be obtained.
  }
\label{fig:3examples}
\end{figure}

Whether or not the reputation values $R_{i}(T)$ converge to positive stationary values very much depends on the topology of the network expressed by the adjacency matrix $\mathcal{A}$, as illustrated in Figure~\ref{fig:3examples}.
Specifically, if an agent has no incoming links that boost its reputation, $R_{i}(t)$ will go to zero.
Therefore, even if this agent has an outgoing link to other agents $j$, it cannot boost their reputation.
Non-trivial solutions depend on the existence of \emph{cycles}, which are formally defined as subgraphs with a closed path from every node in the subgraph back to itself.
The shortest possible cycle involves two agents, $1\to 2 \to 1$.
This maps to \emph{direct reciprocity}: agent 1 boosts the reputation of agent 2 and vice versa.
Cycles of length 3 map to \emph{indirect reciprocity}, for example $1\to 2 \to 3 \to 1$.
In this case, there is \emph{no} direct reciprocity between any two agents, but all of them benefit regarding their reputation because they are part of the cycle.
In order to obtain a non-trivial reputation, an agent not necessarily has to be part of a cycle, but it has to be connected to a cycle.

\section{Dynamics \emph{of} the social network}
\label{sec:network-dynamics}

\subsection{Entry and exit dynamics}
\label{sec:time-evolution}

We now have all elements in place to model the entry and exit dynamics of agents in the OSN.
At each time step $T$, agents evaluate their benefits and costs according to Eqs. \eqref{eq:benefit}, \eqref{eq:cost}.
This is based on their relative reputation $r_{i}(T)$ which has reached a stationary value at time $T$, according to Eqn. \eqref{eq:singlenode}.
They then make a (deterministic) decision to either stay or leave the OSN, according to Eqn. \eqref{eq:choice}.

Hence, at every time $T$, a number $N^{\mathrm{ex}}(T)<N$ of agents will leave the network.
To compensate for this, we assume that the \emph{same} number of new agents will enter the network at the same time, i.e., $N=$const. all the time.
One may argue that this is at odds with our research question, namely to model how cascades of users leaving impact the robustness of the OSN.
But as the empirical case study of the collapse of the OSN \texttt{Friendster} has demonstrated \citep{Garcia2013}, this collapse was \emph{not} due to the fact that no new users entered.
Instead, they became \emph{less integrated} into the social network.
Signs for this trend became already visible when \texttt{Friendster} had about 80 million users.
After that, it still grew up to 113 million users, until it collapsed.
So, the problem of the robustness of an OSN cannot be trivially reduced to the (wrong) assumption that there is a lack of new users entering.

Therefore we have to address the question of how, \emph{despite entering of new users}, large drop-out
 cascades become increasingly likely.
To measure the size of the \emph{drop-out
 cascades}, we will monitor $N^{\mathrm{ex}}(T)$ over time.
If this number is consistently large, it becomes evident that even with a large entry rate, new agents cannot substantially stabilize the OSN, hence its robustness is lost.
We further need to study how new agents will be \emph{integrated} in the OSN. 
If at any time $T$ a varying number of $N^{\mathrm{ex}}(T)$ agents \emph{enter}, we have to model how they are linked to the network, to become members of the OSN.
We assume that new agents do not have complete knowledge of the network; therefore, to start with, they form \emph{random connections} to a (varying) number of members.
Precisely, as in random graphs, new agents create directed links to established agents with a small probability $p$.
Thus, their \emph{expected number} of links is roughly $Np$.

Because agents leaving delete all their links and agents randomly entering  create links, the topology of the network continuously changes at the time scale $T$.
To ensure that the evolution also continues if \emph{no} agent has decided to leave, in this case, we randomly pick one of the agents with the lowest relative reputation, to replace it with one new agent.
To measure how well new agents become integrated into the OSN, we monitor the mean coreness $\mean{k}(T)$, Eqn.~\eqref{eq:2}, over time $T$.
Large values indicate that most agents belong to the core, small values instead that most agents belong to the periphery.

\subsection{Results of computer simulations}
\label{sec:results}

In the following, we discuss the simulation results for a network of fixed size, $N=20$.
Further we use fixed parameters $\gamma=0.1$, $b=1$, $c_{0}=0.45$, $c_{1}=0.05$, $p=0.05$.
For a discussion of parameter dependencies and optimal values, see Section~\ref{sec:lifesp-before-breakd}.

\begin{figure}[htbp]
	\hfill\includegraphics[width=0.21\textwidth]{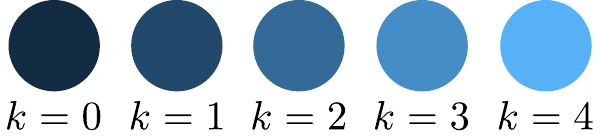}
	
	\vspace{1em}
	\begin{subfigure}{0.24\textwidth}\centering
    \includegraphics[width=\textwidth]{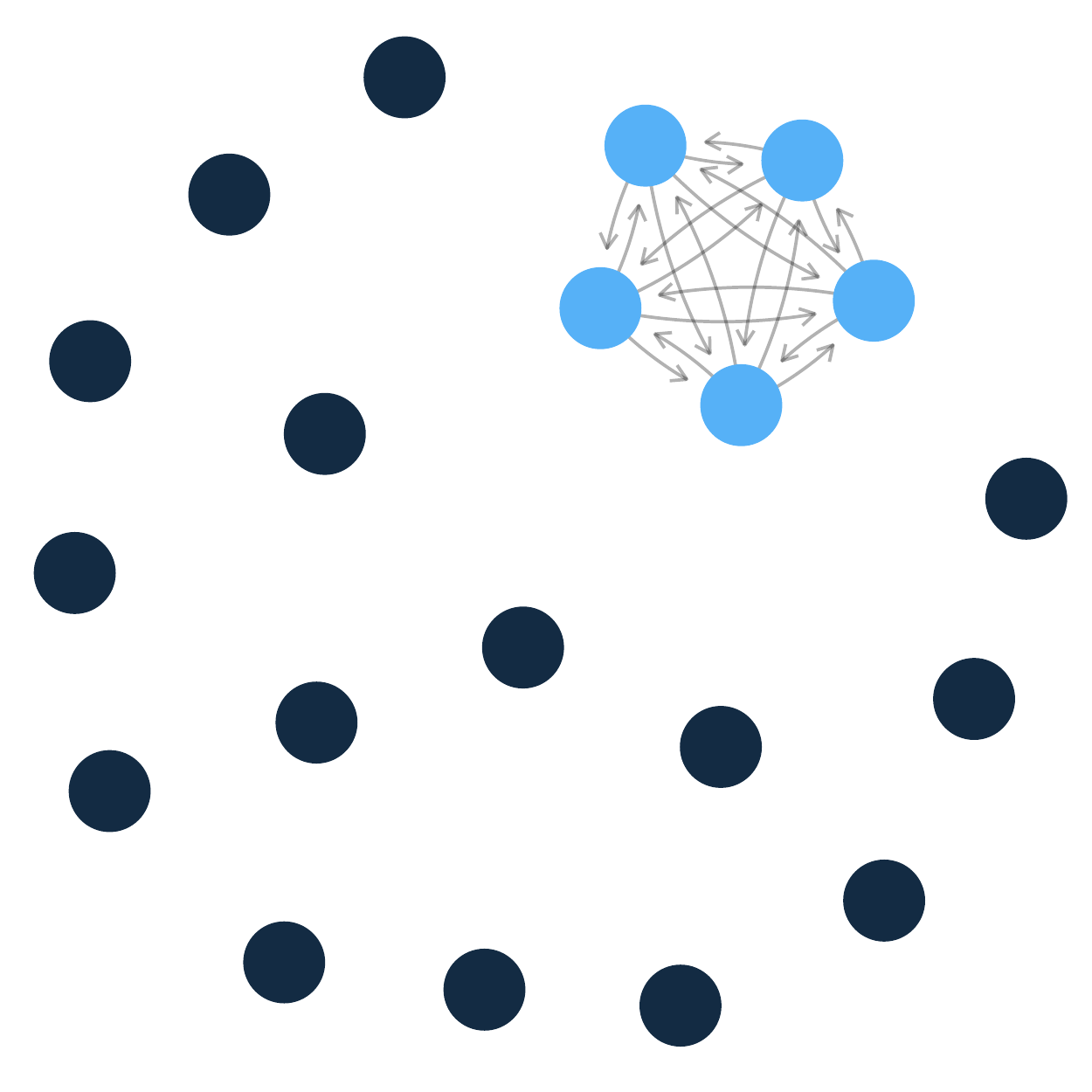}\caption{$T=0$, $\mean{k}=1$}
    \end{subfigure}\hfill
    \begin{subfigure}{0.24\textwidth}\centering
    \includegraphics[width=\textwidth]{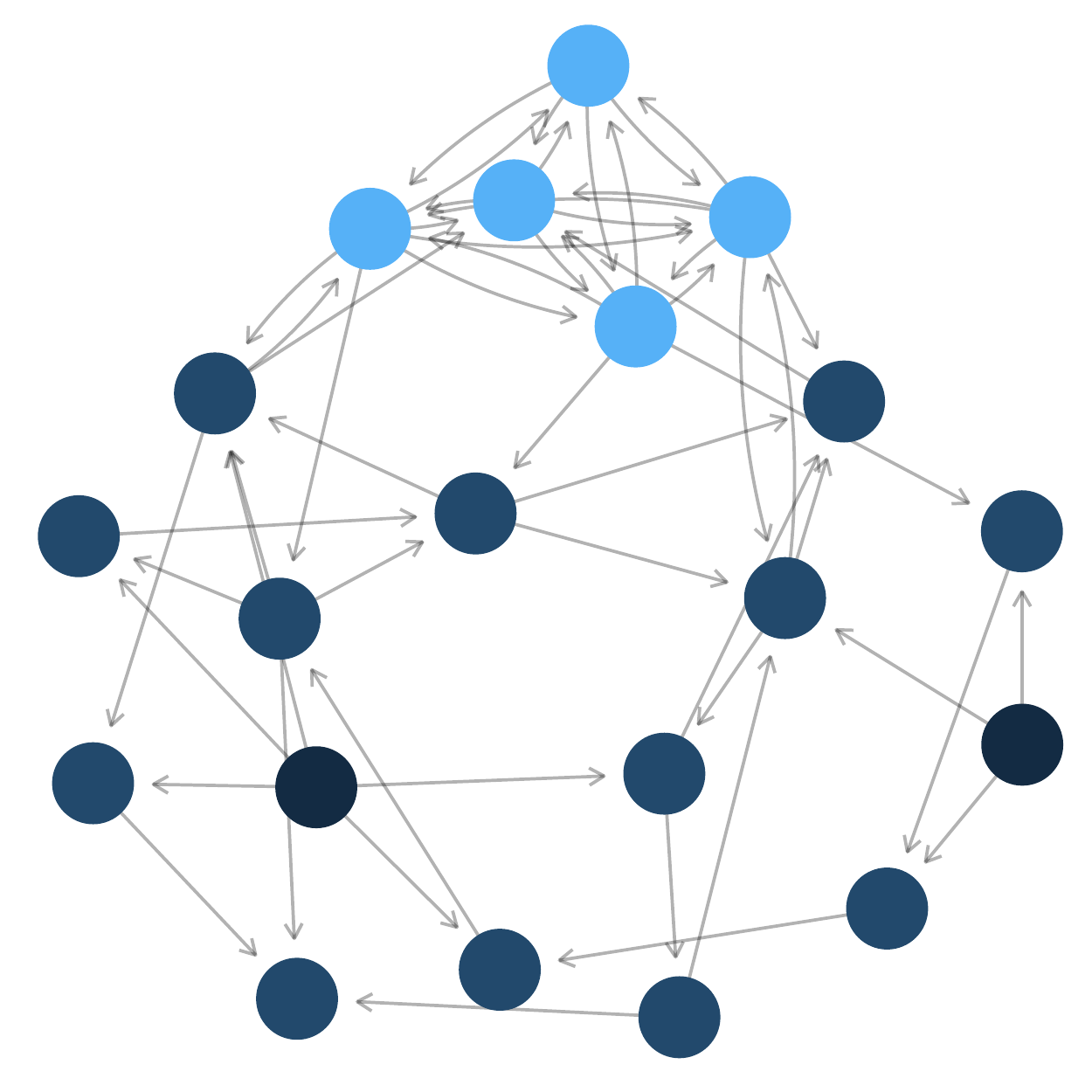}\caption{$T=50$, $\mean{k}=1.65$}
    \end{subfigure}\hfill
    \begin{subfigure}{0.24\textwidth}\centering
    \includegraphics[width=\textwidth]{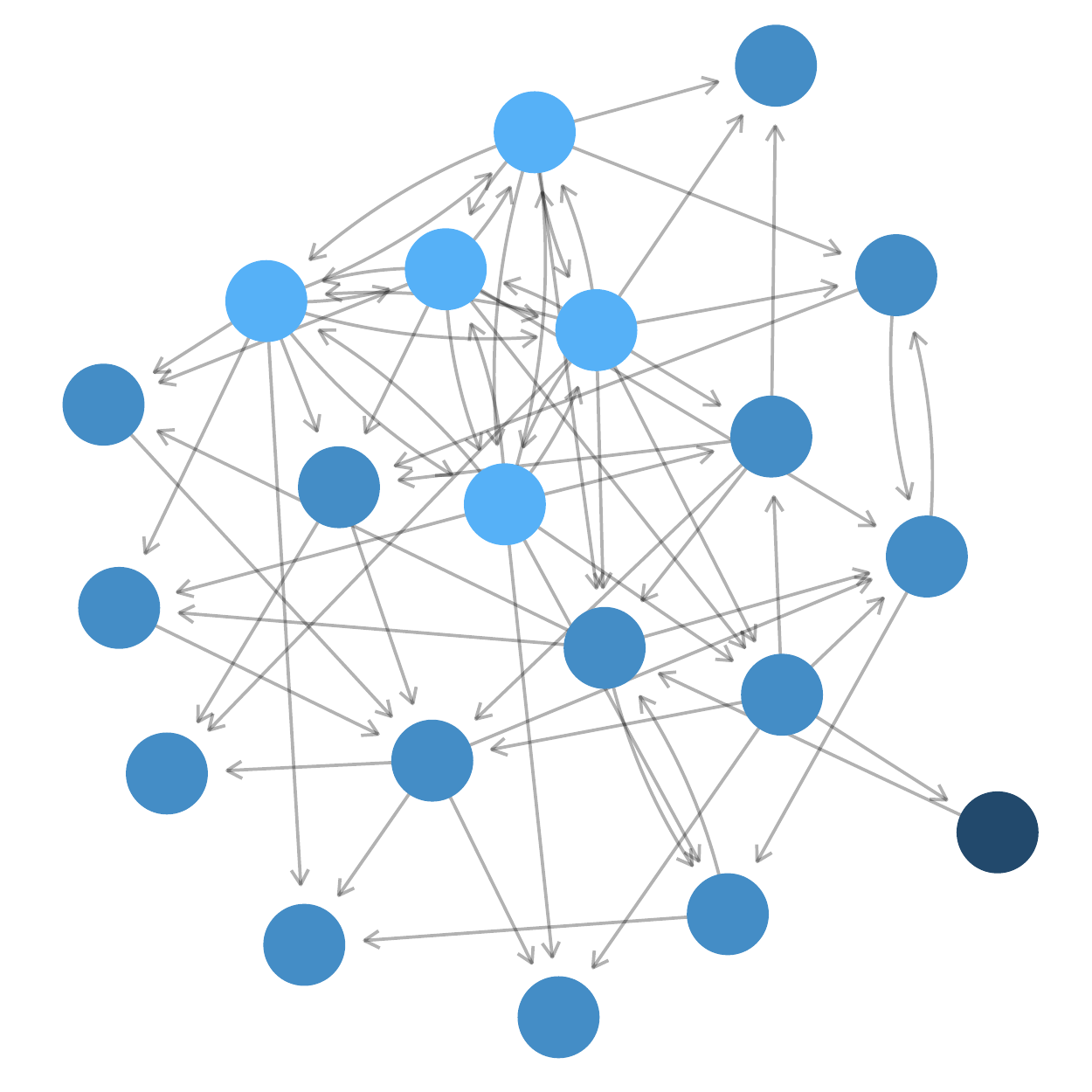}\caption{$T=501$, $\mean{k}=3.15$}
    \end{subfigure}\hfill
    \begin{subfigure}{0.24\textwidth}\centering
    \includegraphics[width=\textwidth]{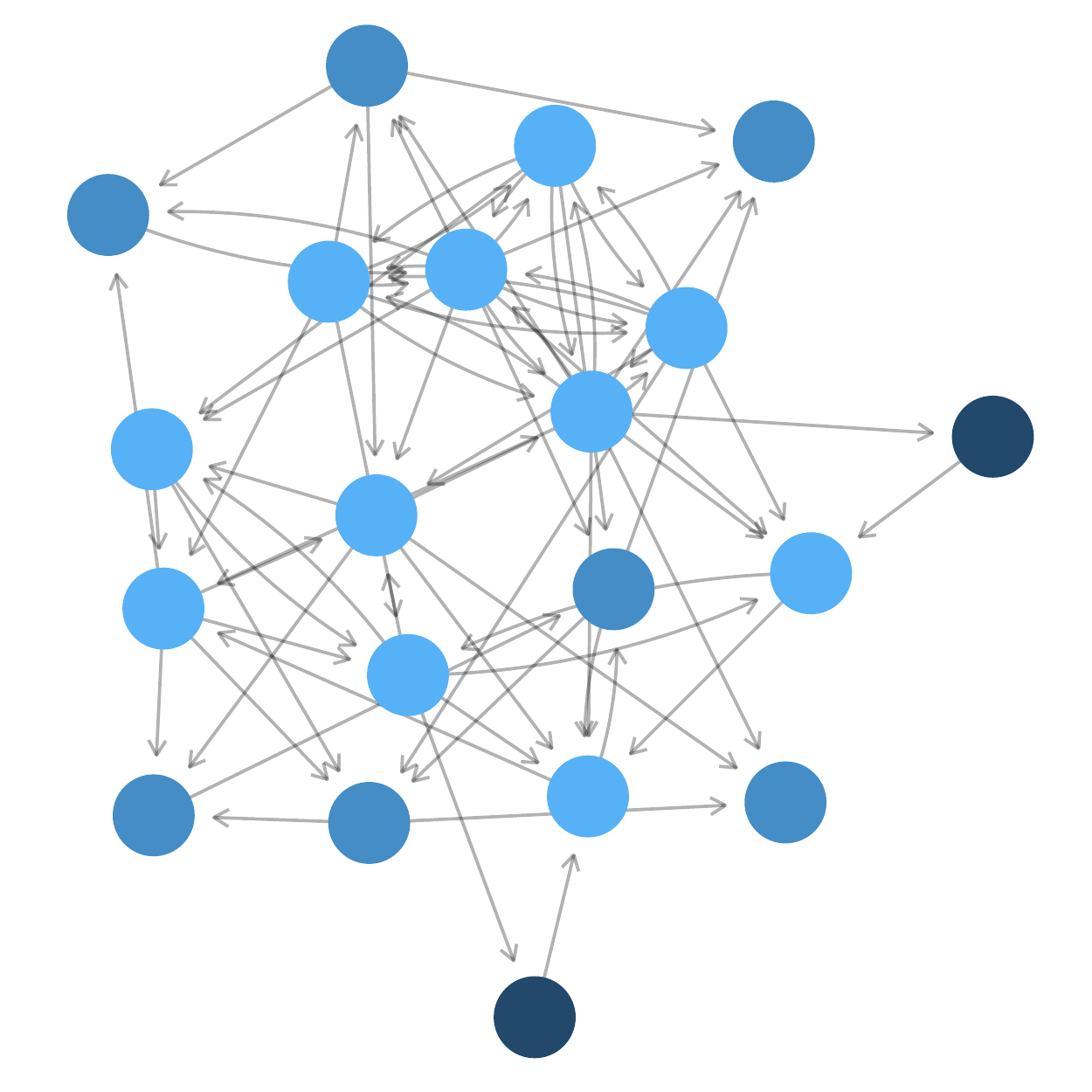}\caption{$T=3001$, $\mean{k}=3.35$}
    \end{subfigure}
    
    \vspace{1em}
    \begin{subfigure}{0.24\textwidth}\centering
    \includegraphics[width=\textwidth]{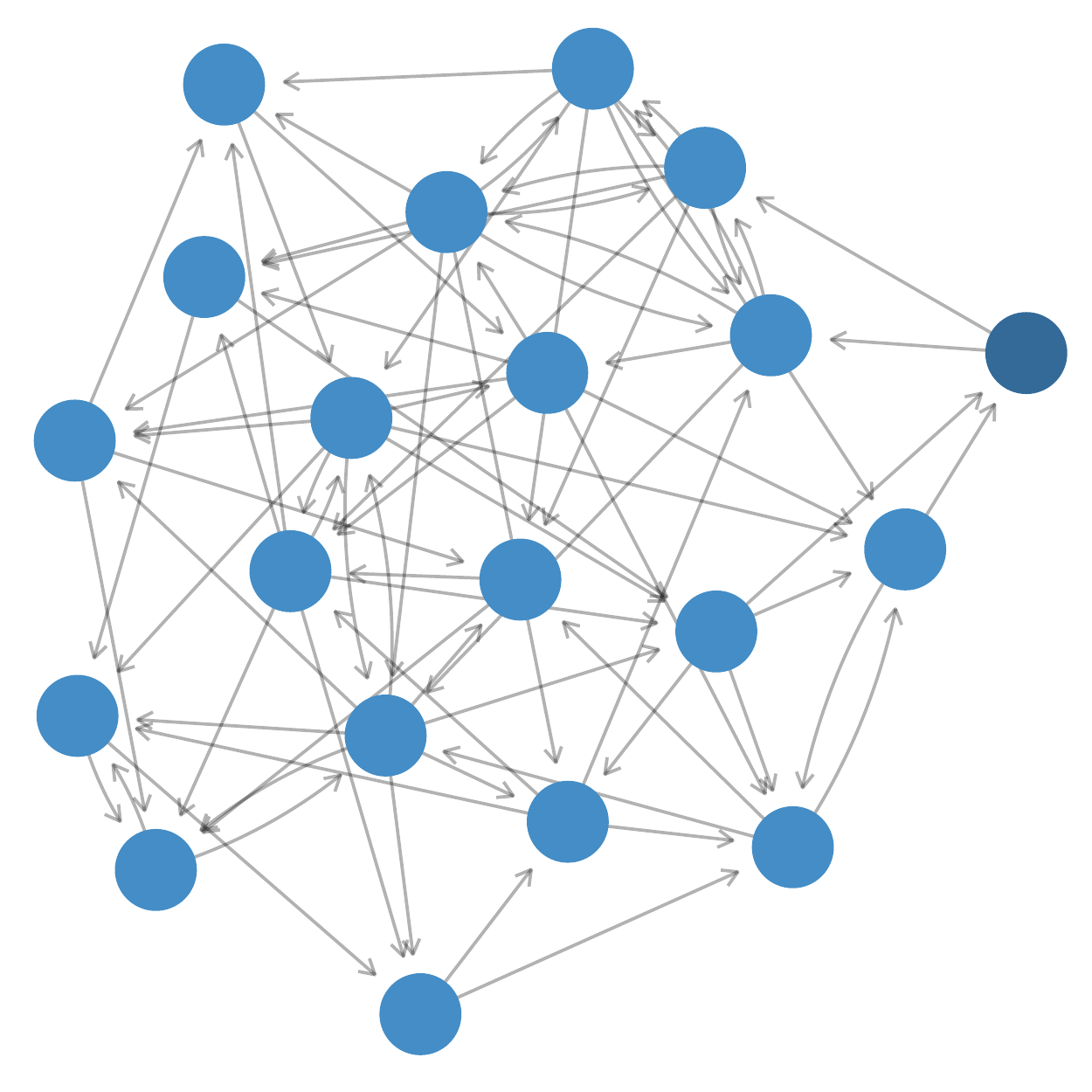}\caption{$T=5350$, $\mean{k}=2.95$}
    \end{subfigure}\hfill
    \begin{subfigure}{0.24\textwidth}\centering
    \includegraphics[width=\textwidth]{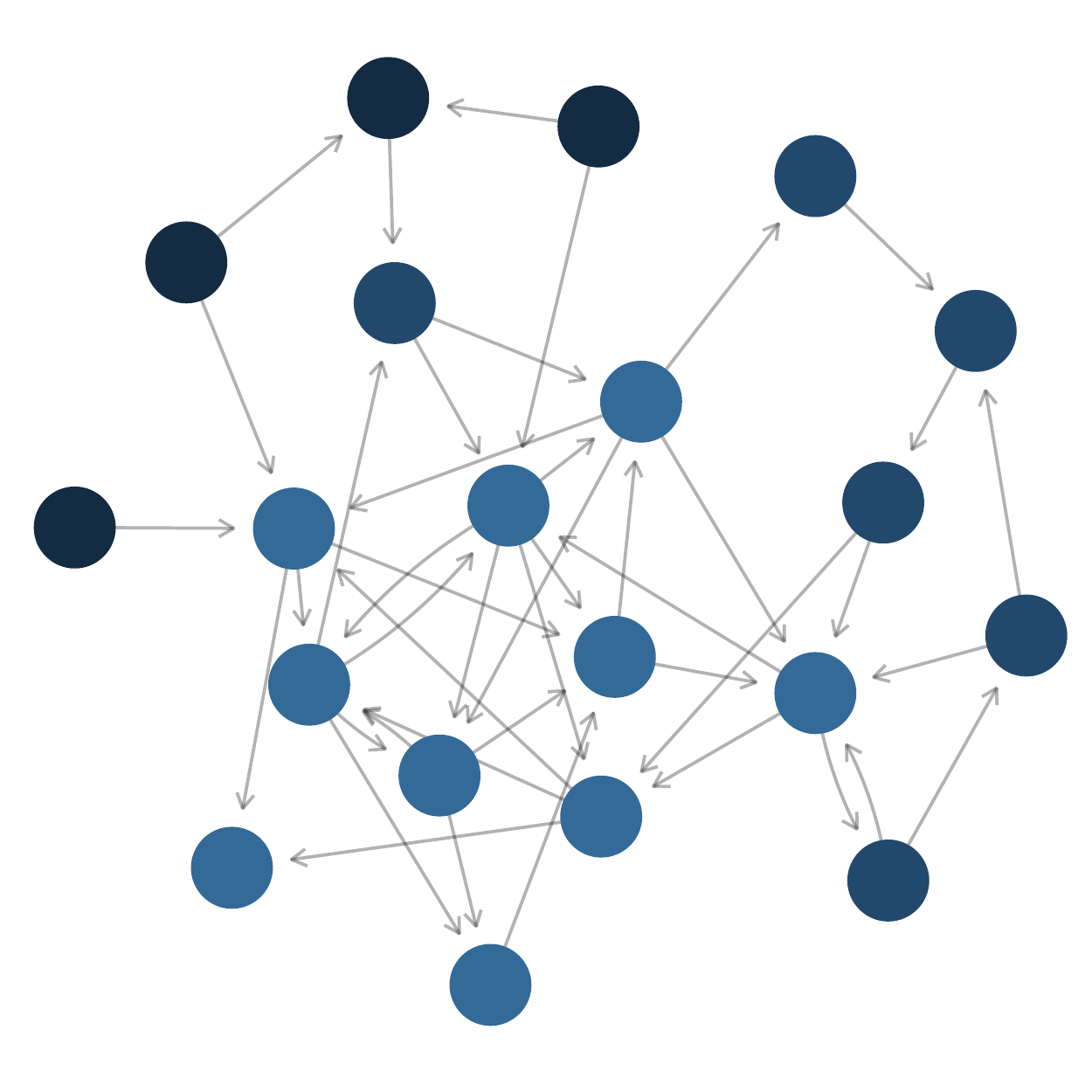}\caption{$T=5353$, $\mean{k}=1.3$}
    \end{subfigure}\hfill
    \begin{subfigure}{0.24\textwidth}\centering
    \includegraphics[width=\textwidth]{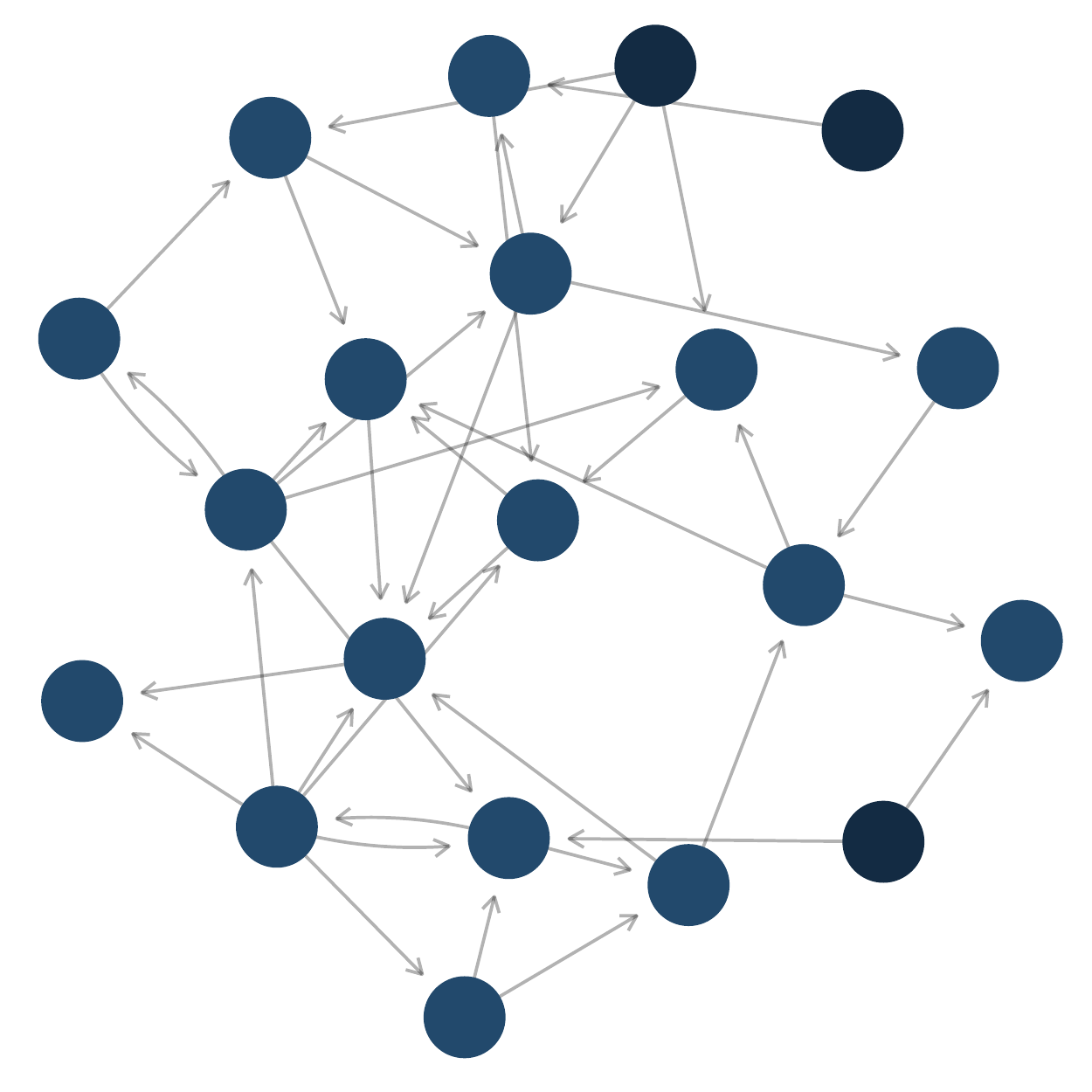}\caption{$T=5360$, $\mean{k}=0.85$}
    \end{subfigure}\hfill
    \begin{subfigure}{0.24\textwidth}\centering
    \includegraphics[width=\textwidth]{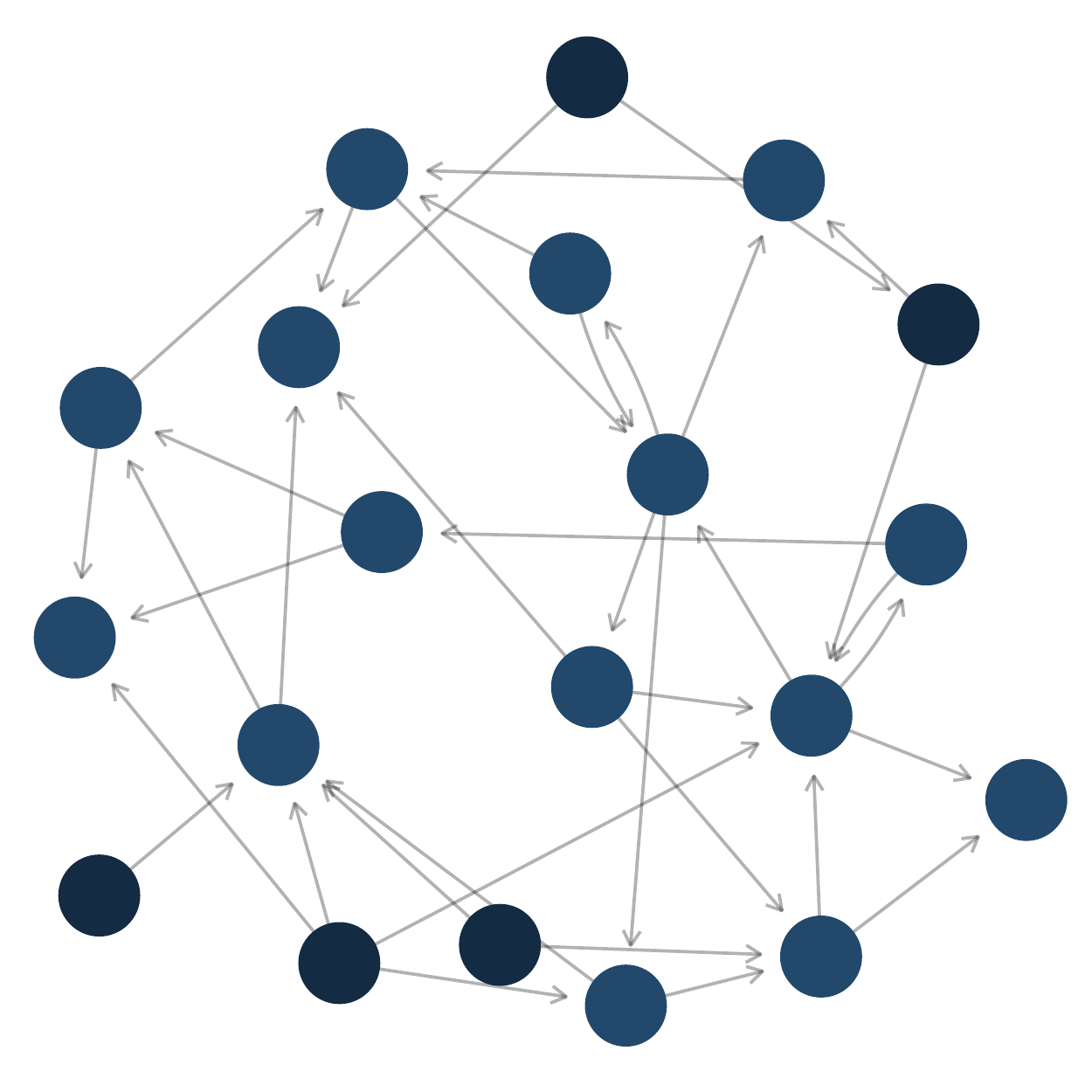}\caption{$T=7000$, $\mean{k}=0.75$}
    \end{subfigure}
    \caption[]{Some instances from the graph evolution of a $20$-nodes network.
}
  \label{fig:20graphevo}
\end{figure}

\begin{figure}[htbp]
  \centering
  \includegraphics[width=.65\textwidth]{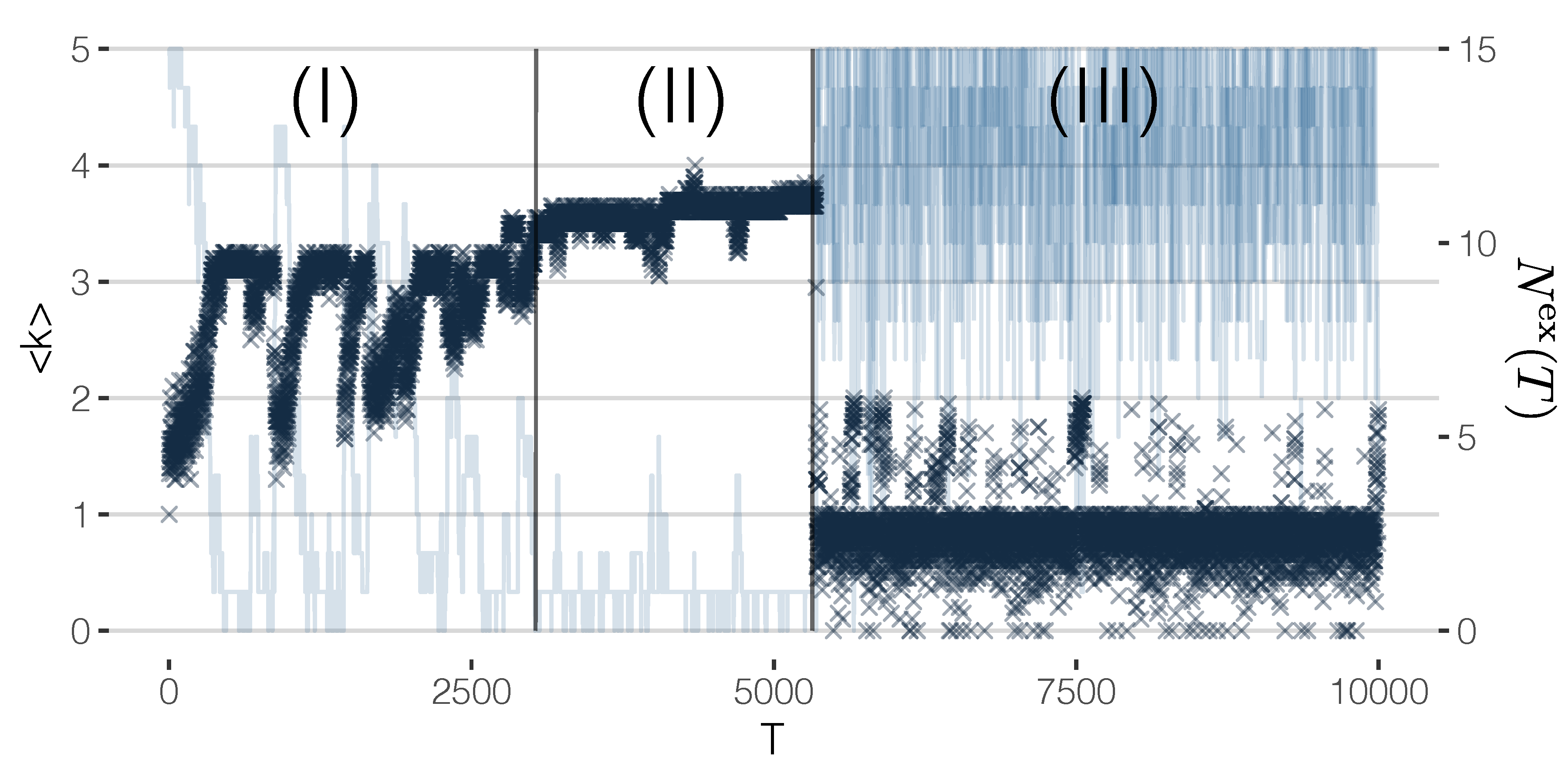}
   \caption[Mean coreness and number of rewired nodes for $20$-nodes graph]{Evolution of mean
    coreness and number of rewired nodes for each time step in a $20$-nodes network. 3 regions can
    be identified: (I) Build up, (II) Metastable state, (III) Breakdown.}
  \label{fig:20evo}
\end{figure}

To initialize our simulations of the network dynamics, we assume that at time $T=0$, 5 out of 20 agents initially form a fully connected cluster, as shown in Figure~\ref{fig:20graphevo}(a).
This ensures that these five agents have a non-zero reputation at $T=1$ and thus will not leave the OSN.
The remaining 15 agents with reputation zero, however, will be replaced by new agents that randomly create links to the agents in the network.
This way, at $T=50$ already a realistic network structure with a \emph{core}, a \emph{periphery}, different $k$-\emph{shells} and a few isolated agents emerges, as shown in Figure~\ref{fig:20graphevo}(b).
Figure~\ref{fig:20graphevo} displays further snapshots of the network evolution, while 
the corresponding systemic variables to monitor the dynamics, namely the mean coreness, $\mean{k}(T)$, and the number of agents leaving, $N^{\mathrm{ex}}(T)$, are shown in Figure~\ref{fig:20evo}.
From the latter, we can clearly identify three different phases of network evolution. 

\paragraph{(I) Build up phase. \ }

In this initial phase, as already mentioned, the network establishes its characteristic topology. 
Most agents become tightly integrated into the network, as also visible from Figures~\ref{fig:20graphevo}(b,c).
Because of this, the mean coreness quickly increases, while the number of agents leaving decreases, but both variables show considerable fluctuations.

\paragraph{(II) Metastable phase. \ } 

After agents have become well connected to the core, they tend to have higher benefits than costs.
If no agent would leave the OSN, we choose one of the agents with the lowest reputation to leave, to keep the network dynamics going. 
Hence, $N^{\mathrm{ex}}(T)=1$ or very low, for most of the time, while $\mean{k}$ only slightly fluctuates.

Still, the status of the OSN is not stable but only \emph{metastable}, because of the slow dynamics that is illustrated by means of Figures~\ref{fig:20graphevo}(e,f).
Agents that were earlier part of the periphery have now become part of the core, this way \emph{decreasing} the size of the periphery.
In fact, the smaller the periphery, the more likely the formation of new links to the core.
The probability that a new agent $i$ becomes part of the core $Q$ with size $\abs{Q}$ is given as:
\begin{equation}
  \label{eq:prob-core}
  P(i\in Q)\geq {\abs{Q}\choose k_{\mathrm{max}}^{-}}\; p^{k^{-}_{\mathrm{max}}} 
  \cdot
  {\abs{Q}\choose k_{\mathrm{max}}^{+}}\; p^{k^{+}_{\mathrm{max}}}
\end{equation}
where $k_{\mathrm{max}}^{-}$,  $k_{\mathrm{max}}^{+}$ are the values for the in-degree and the out-degree coreness of the agents in the \emph{core}.
The two r.h.s terms stand for the probability of creating and of receiving links from the
core, where $p$ is the probability for an incoming agent to create a new link.
$P(i\in Q)$ is indeed increasing with the size of the core, $\abs{Q}$~\cite{Luczak1991}.

\paragraph{(III) Breakdown phase. \ }

The slow dynamics during phase (II) leads to a point where agents from the outer shells of the in-degree core receive a higher reputation than agents in the core.
If no agent decides to leave the OSN, in this situation, an agent from the core is chosen to be removed, because of the lower reputation. 
This then triggers whole \emph{cascades} of agents leaving, because the drop-out of a core agent abruptly decreases the reputation of other agents in the core and the outer shells.
The transition from phase (II) to phase (III) can be seen by the \emph{increasing} number of agents leaving, while the mean coreness steadily \emph{decreases}.

Once the core has been destroyed, the OSN has no ability to recover because most agents are replaced at each time step.
Nearly all links from the newly entering agents will be to agents from the periphery; thus, the probability of forming a new core is extremely low. The breakdown phase (III) can be characterized not only by the rather low mean coreness and the large number of entries and exits, but also by the much larger fluctuations of both values.

\section{Improving Robustness}
\label{sec:improving-robustness}

\subsection{Network interventions}
\label{sec:netw-interv}

The simulation results shown in Figure~\ref{fig:20evo} make it very clear what we mean by improving robustness: to prevent the \emph{complete breakdown} of the OSN.
This does not imply to prevent cascades, which can always happen in response to agents leaving the OSN. 
But we argue that a social network is \emph{robust} if the decision of agents to leave the OSN will not trigger large cascades of leaving agents that destroy the whole core. 

This requires us to influence agents in the OSN such that they decide \emph{not} to leave the network.
The trivial solution would be to reduce the costs of \emph{all} agents to a level that always guarantees a positive utility or to increase the benefits in the same manner.
A much smarter solution, however, would focus only on a \emph{few} agents, namely those with the ability to prevent large cascades.
The problem to identify those agents is addressed in research about network controllability \citep{Liu2011,Zhang2016}, which is related to control theory.
The method assigns a \emph{control signal}, i.e., an incentive to stay or to leave, to the identified agents with the most influence on the network dynamics \citep{zhang2019control}, which are called \emph{driver nodes}. 
Precisely, this signal is added to the reputation dynamics, Eqn.~\eqref{eq:singlenode}, of the driver nodes.

We will not follow this formal procedure in our paper for several reasons.
The most important one is the continuous evolution of the network topology, which is not considered in the network controllability approach.
It would require us to redo the identification of the driver nodes and the assignment of control signals at every time step $T$.
Further, in our context of users leaving an OSN, these control signals are difficult to interpret because they change the \emph{reputation} dynamics.
Our intention instead is to influence the \emph{decisions} of the agents, Eqn.~\eqref{eq:choice}, i.e. to apply control signals to the \emph{costs} of staying in the OSN.
Specifically, we apply two different scenarios to incentivize agents (i) from the periphery, or (ii) from the core. 

\begin{figure}[htbp]
\begin{subfigure}{0.45\textwidth}
     \includegraphics[width=\textwidth]{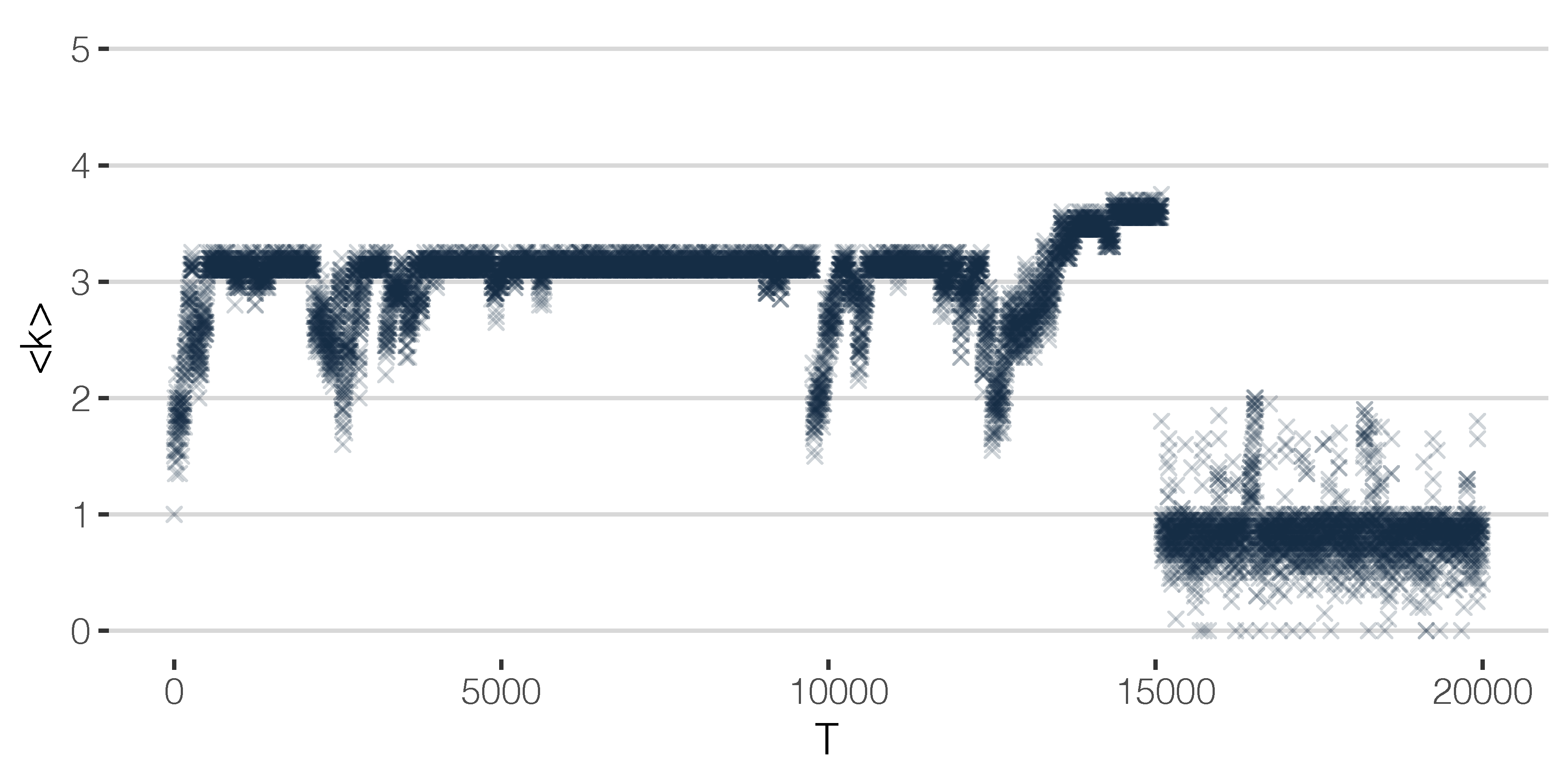}
     \caption{}
     \end{subfigure}\hfill
     \begin{subfigure}{0.45\textwidth} 
     \includegraphics[width=\textwidth]{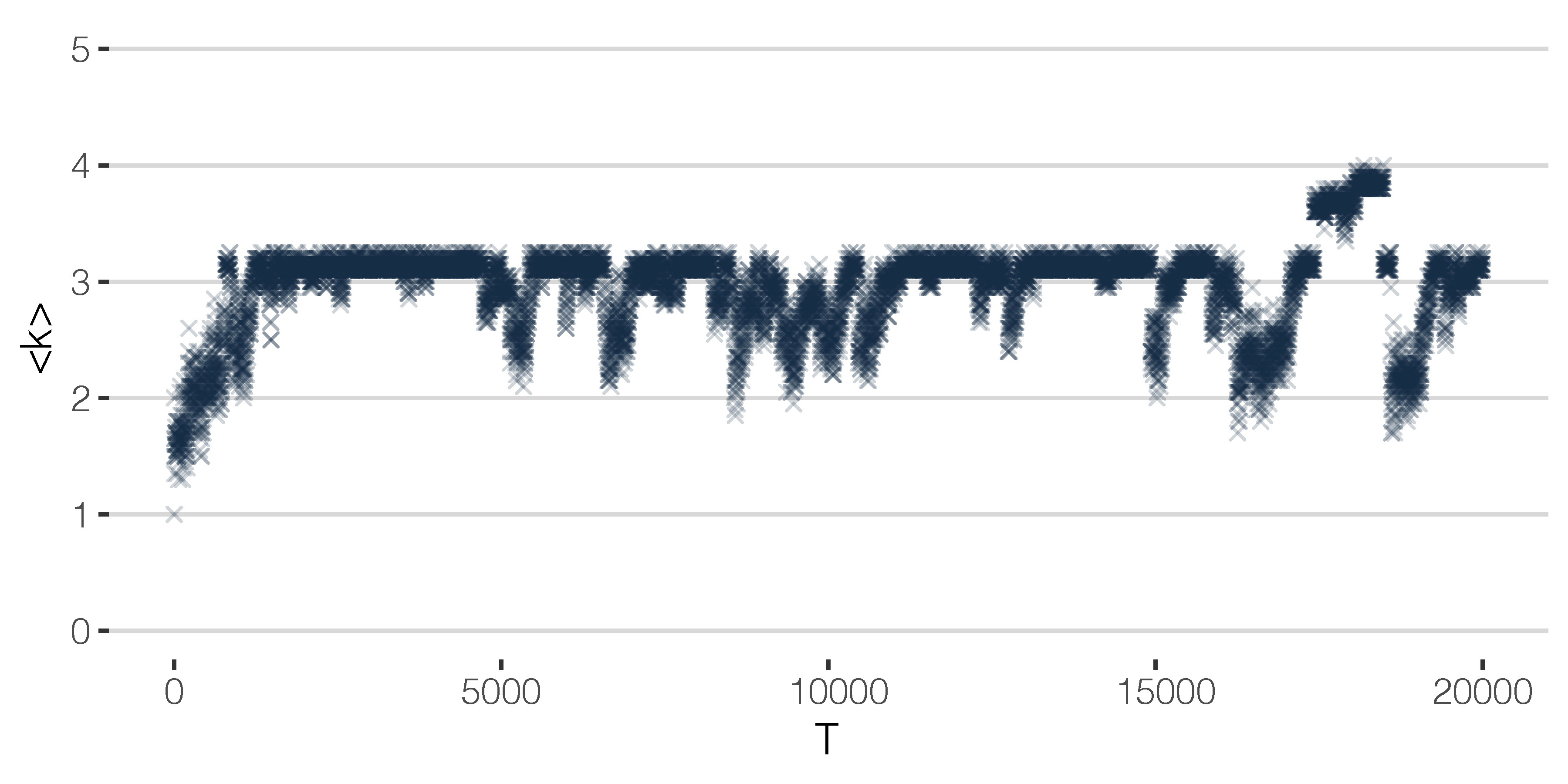}
     \caption{}
     \end{subfigure}
     
   \hspace*{1.5cm}
  \begin{subfigure}{0.33\textwidth}
  \includegraphics[width=\textwidth]{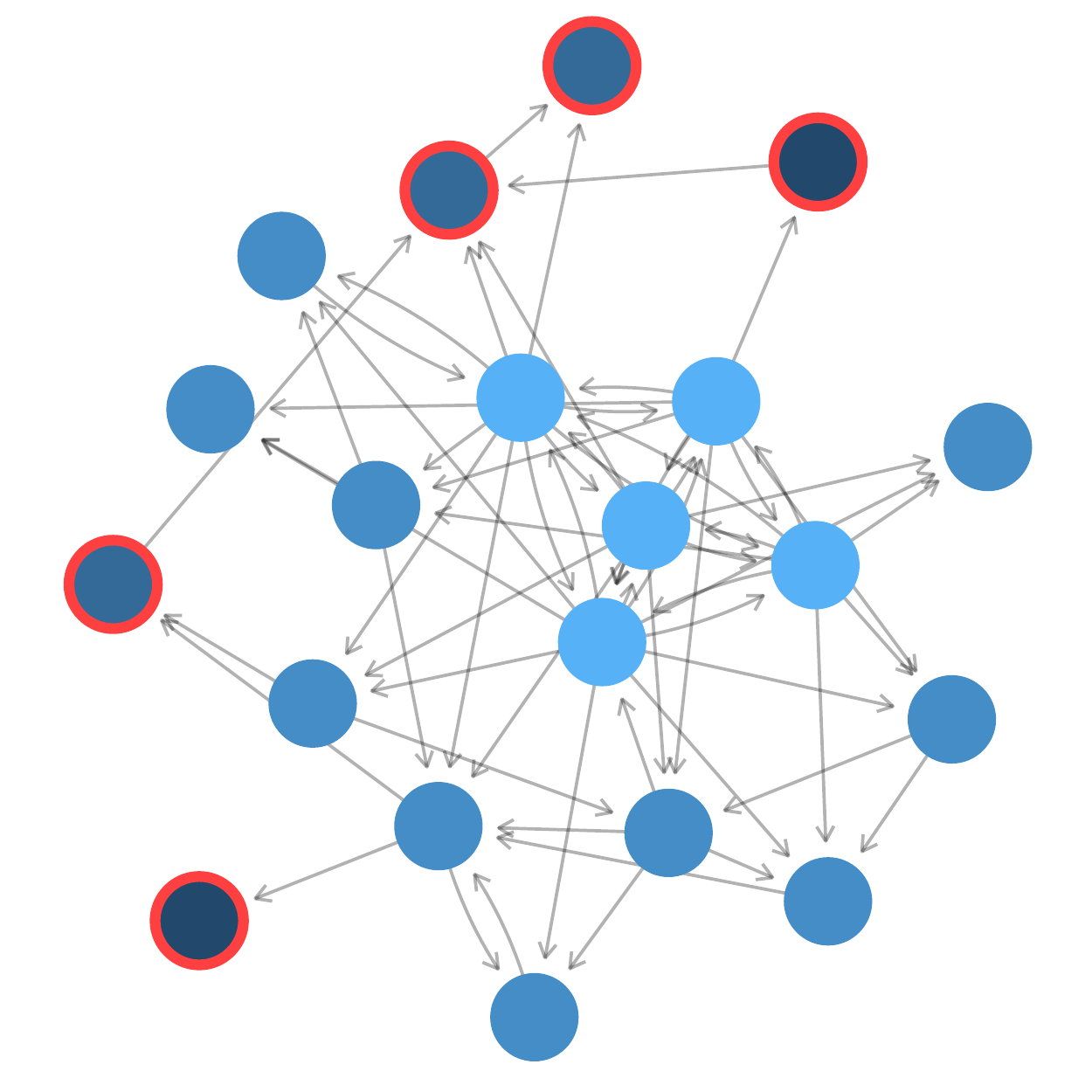}
  	\caption{}
     \end{subfigure}\hfill
     \begin{subfigure}{0.33\textwidth} 
  \includegraphics[width=\textwidth]{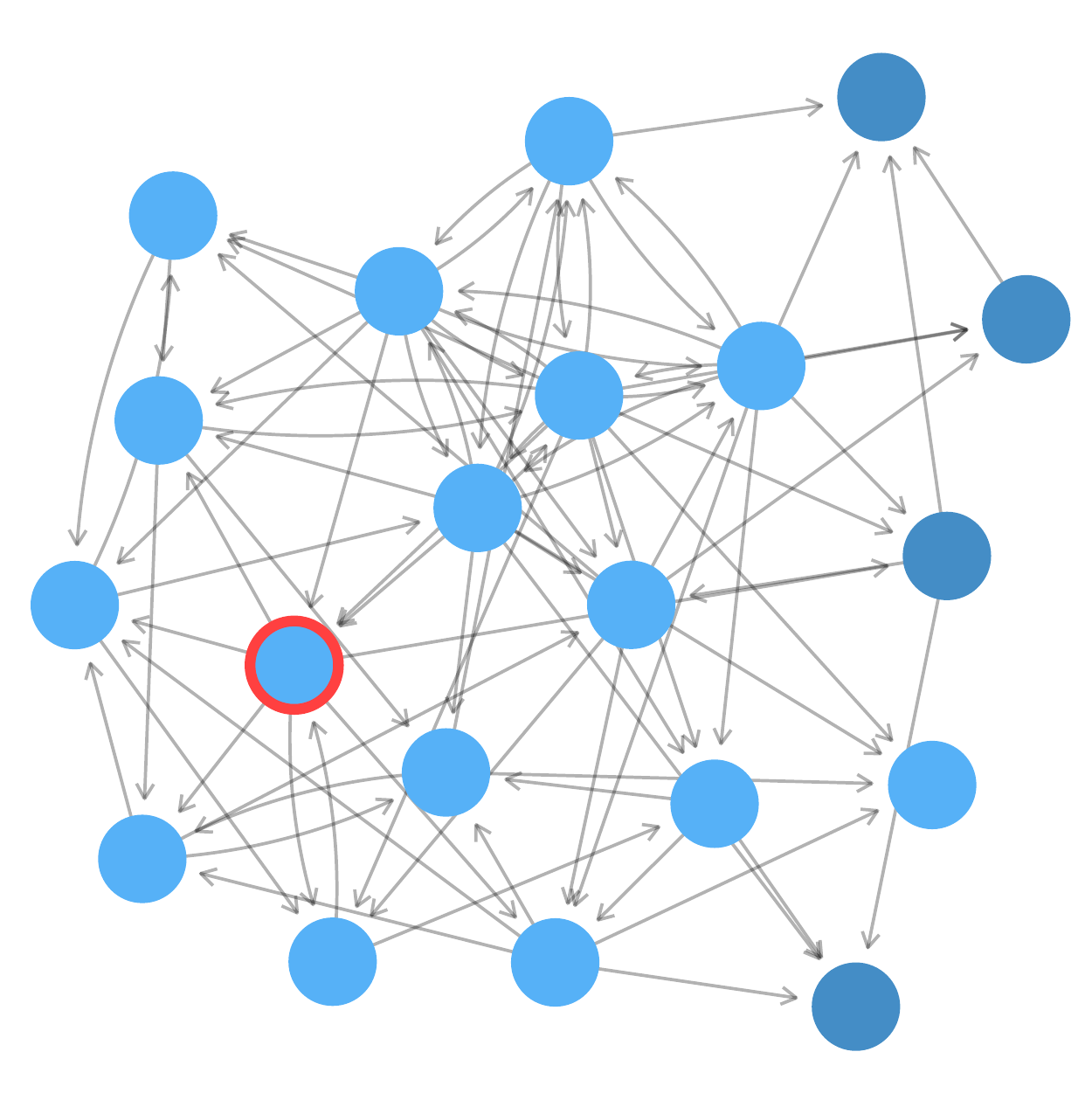}
       \caption{}
     \end{subfigure}
     \caption{Results of network interventions.  (left) control of many peripheral agents, (right) control of one agent close to the core. (a,b) show the mean coreness $\mean{k}$ and the number of agents leaving, $N^{\mathrm{ex}}$ over time $T$, to be compared to Figure~\ref{fig:20evo}. (c,d) Snapshots of the network at a particular time $T$, when the cost $c_{0}$ of the agents circled in \emph{red} is adjusted. }
  \label{fig:control-net}
\end{figure}

The first scenario is motivated by our insight that large cascades are caused by the \emph{disappearing periphery}.
Therefore, a straightforward intervention is to choose agents with a low reputation from the periphery as drivers.
These are incentivized to \emph{stay} in the OSN, i.e., their costs are \emph{reduced} such that their utility is increased and they decide \emph{not} to leave. 
The second scenario is to choose agents close to the core, i.e., from its first outer shells, as drivers.
These are incentivized to \emph{leave} the OSN, i.e. their costs are \emph{increased} such that they decide to \emph{not stay}.
This more subtle scenario is motivated by the insight that agents that are only close to the core will \emph{not} trigger large cascades if they leave.
But if they leave, they considerably reduce the reputation of their closest neighbors, this way \emph{increasing} the size of the
periphery.
The results of these two scenarios are illustrated in Figure~\ref{fig:control-net}.

Specifically, in scenario (i), we identify at each time step $T$ all agents from the periphery, i.e., with a coreness value $k_{i}=1$.
Their cost $c_{0}$ is then reduced by 10 percent, i.e. to $\hat{c}_{0}=0.9c_{0}$.
As Figure~\ref{fig:control-net}(a) demonstrates, this scenario can only delay the complete breakdown (in comparison to Figure~\ref{fig:20evo} without any interventions).
But it cannot completely prevent large drop-out
 cascade, because the build-up of a large core that eventually gets destroyed is only delayed.

In scenario (ii), on the other hand, we are able to achieve the goal of preventing a complete breakdown.
This scenario has remarkable differences to scenario (i): We only incentivize \emph{one} agent, instead of many, and we choose this agent from the vicinity of the \emph{core} instead from the periphery.
Precisely, we choose the agent from the first outer shell identified by means of the directed $k$-core decomposition, i.e., $k_{i}=k_{\mathrm{max}}-1$. 
This agent is enforced to leave by increasing its cost by 10 percent, i.e. to  $\hat{c}_{0}=1.1 c_{0}$. 

As shown in Figure~\ref{fig:control-net}(b), this scenario considerably improves the robustness of the network, as witnessed by the average coreness.
At the same time, because one agent is chosen for control from the beginning, we also observe that the build-up phase (I) is extended in comparison to the case of no control (see Figure~\ref{fig:20evo}).
But phase (II), which was called metastable before, is now considerably extended.
We still notice small cascades, but no complete breakdown, i.e., the metastable phase has become a \emph{quasistable} one.

\subsection{Life-time before breakdown}
\label{sec:lifesp-before-breakd}

The above simulations are both interesting and counter-intuitive because controlling one agent close to the core leads to much better results than controlling many agents from the periphery.
We, therefore, continue with a more refined discussion of the peripheral control.
As shown, this kind of network intervention increases the time before the breakdown, but cannot completely prevent it.
To further quantify this dynamics, we use the \emph{life-time} $\Omega_{Q}$ of the core $Q$ (measured in network time $T$) as an additional systemic variable \citep{mavrodiev2019}.
As Figure~\ref{fig:control-net}(a) illustrates, for scenario (i)
the value of $\Omega_{Q}$ can be clearly obtained from the simulations because of the sharp transition toward the breakdown of the OSN. 
For scenario (ii), obviously $\Omega_{Q}\to \infty$ as Figure~\ref{fig:control-net}(b) shows.

\begin{figure}
  \centering
  \includegraphics[width=0.6\textwidth]{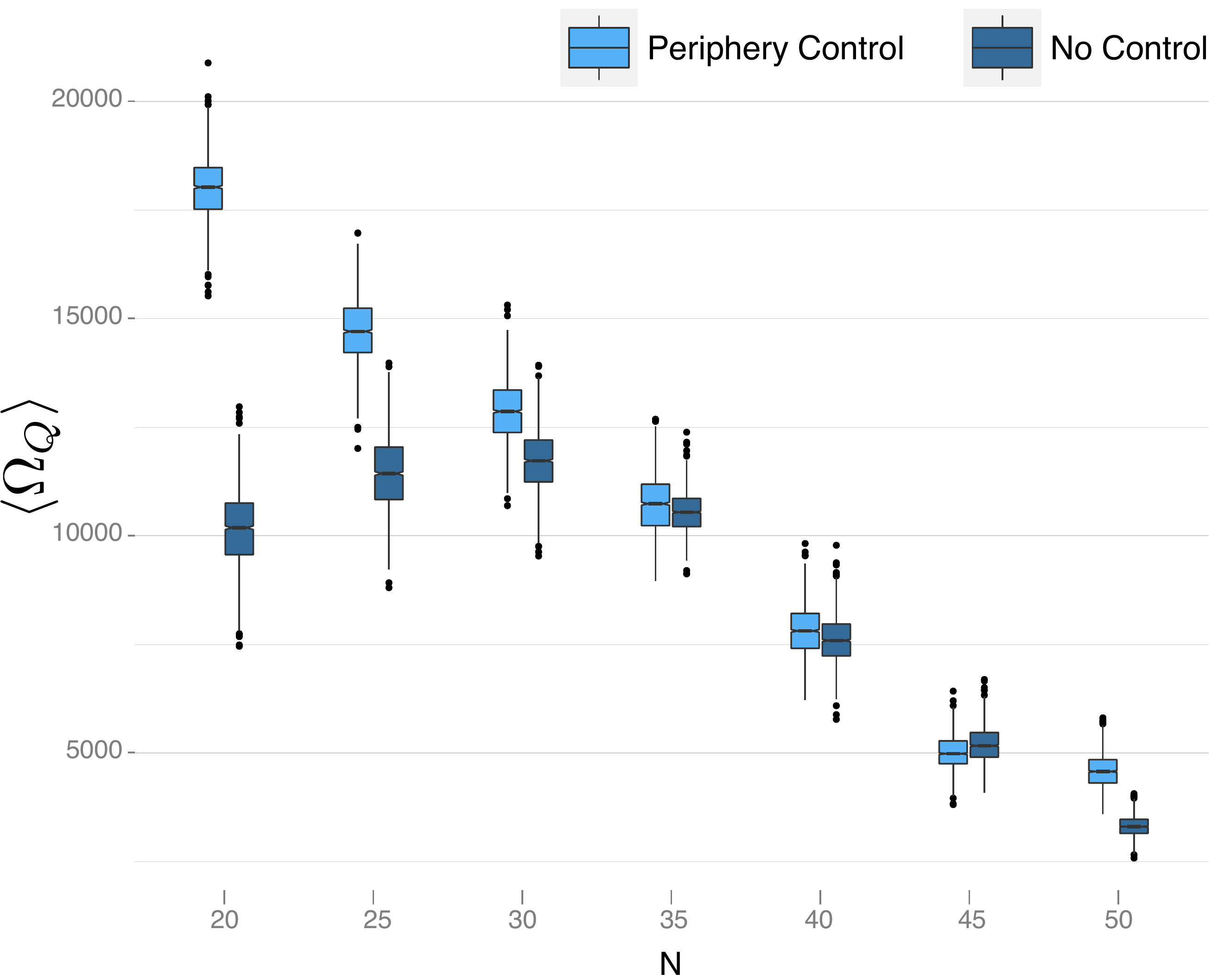}
  \caption[Control periphery bootstrap outcome]{Comparison of different periphery control approaches
    with fixed control signal. The effectiveness of the control method without adapting the signal
    to the size of the network decreases with size. In the figure are plotted bootstrap samples for $\mean{\Omega_{Q}}$ 
    obtained from 100 simulations for each network size and each strategy. The control signal used is $u=-0.05$.}
  \label{fig:bootcontrol}
\end{figure}

We are interested in comparing the life-times of the core for peripheral control and without control (also shown in Figure~\ref{fig:20evo}).
Because $\Omega_{Q}$ changes considerably for different simulations, we use the average life-time $\mean{\Omega_{Q}}$ taken from 100 independent runs with the same setup.
We further have to consider that  $\mean{\Omega_{Q}}$ depends on other system parameters, notably the system size $N$.
We, therefore, vary $N$ for simulations with peripheral control and without control, keeping all other parameters the same. 
The results are shown in Figure~\ref{fig:bootcontrol}, from which we can deduce some interesting insights.

First, we note that for \emph{small} networks ($N<30$), our peripheral control strategy works very well.
The life-times increased considerably in comparison to the no-control reference case.
Secondly, we observe that this advantage becomes smaller if the network size increases.
For networks larger than $N=30$, there is almost no difference in life-times between the peripheral control and the no-control case.
Further, for $N>30$ in both cases, the life-time decreases almost linearly with the increasing network size.

The latter observation can be explained from the fact that, with increasing network size $N$, the network becomes much denser.
We recall that links between agents are formed such that new agents entering the OSN create links to established agents with a \emph{fixed probability}, $p$.
The average number of links per agent is thus $Np$, i.e., it increases linearly with $N$.
The denser the network, the larger the core and the smaller the periphery.
In line with our above discussion, this means less robustness of the network, i.e., the breakdown occurs earlier in time.

The non-monotonous dependence of  $\mean{\Omega_{Q}}$ on the network size, for the \emph{no-control} case, results from the fact that the model parameters are not completely independent.
This fact is also obvious from Eqn.~\eqref{eq:1}. 
Instead, it was already pointed out \citep{mavrodiev2019} that there is an \emph{optimal cost level} to maximize the life-time of the network.
This is understandable from our above discussions.
If costs are very low, only very few agents will leave the OSN.
Because of the slow dynamics described in phase (II), these agents will, at some point, reach a reputation large enough to compare to the core, and hence the core agents will leave.
An intermediate cost level, on the other hand, makes sure that this evolution does not take place, or is at least considerably delayed.
The optimal cost level that maximizes the life-time, however, also depends on the other parameters, $b$, $N$, $\gamma$, $p$. 

From  Figure~\ref{fig:bootcontrol}, we can deduce that, for the fixed cost parameters chosen in our simulation, the optimal network size is $N=30$, simply because, for this size, the life-time is maximized (kept all other parameters the same).
Hence, for small networks, $N<30$, the optimal cost level should be \emph{lower} than what was used in the simulation.
Given the suboptimal values, the life-time was also lower for the no-control case.
Remarkably, the life-time in case of peripheral control is not affected by this.
So, we can conclude that, at least for small networks, peripheral control also compensates for not optimal parameter choices.

For larger networks, $N>30$, Figure~\ref{fig:bootcontrol} suggests that there is \emph{no difference} between peripheral control and no control.
But this observation is mainly due to the fact that we have not used the optimal parameters for a given network size $N$.
To further investigate this, we have performed an extensive optimization to determine the optimal values for $c_{0}$ and $\hat{c}_{0}$ for a given $N$.
It then turns out that, with the optimal parameters, the life-times for the peripheral control and no-control cases are no longer the same, but differ \emph{significantly}.

Specifically,  we performed two-samples \emph{t-tests} for the means and
\emph{Wilcoxons-tests} for the medians of bootstrap samples of the average life-times $\mean{\Omega_{Q}}$ obtained from the simulations with and without control.
As the $H_{0}$ hypothesis, we assume that the means of the life-times in both cases are equal and as alternative hypothesis that the life-times are higher in case of peripheral control.
Using always the optimal parameters for both cases, we obtained p-values in the order of $10^{-12}$ for the alternative hypothesis, independent of the network size.
This provides strong evidence for the conclusion that the peripheral control \emph{always} improves the robustness of the network, as measured by the life-time before breakdown.
For small networks, this holds already for arbitrary parameter choices, for large networks only if the optimal parameters are chosen.

In Figure~\ref{fig:bootcontrol}, we also plot the bootstrapped $95\%$ confidence intervals for the average life-time $\mean{\Omega_{Q}}$. 
We note that the size of the confidence interval decreases with $N$.
Hence, for small networks, even optimal parameter values cannot guarantee a minimal variance of $\Omega_{Q}$, and in single
simulations, a breakdown of the network can happen much earlier or later. 

Eventually, we also tested whether reputation differences in the peripheral agents matter for the network intervention.
While the above simulations assumed that \emph{all} peripheral agents are controlled, we also considered that only peripheral agents with \emph{high}, or with \emph{low} reputation are influenced in their costs.
These cases, however, did not generate any remarkable difference with respect to the average life-time.

\section{Conclusions}
\label{sec:conclusions}

After more than 35 years of \emph{understanding} complex systems, there should be foundations enough for \emph{managing} them in a better and more quantitative manner.
Sadly, to know how systems \emph{work} does not already imply also to know how to \emph{influence} them such that more desired system states are obtained.
This holds particularly for socio-economic systems, which are \emph{adaptive}, which means they respond to proposed changes in both intended and unintended ways.
\emph{Systems design} \citep{Schweitzer:2019vp} therefore has to master a difficult balance: on the one hand, systems should be carefully steered towards a wanted development, on the other hand, systems should not be over-regulated, to not lose their ability to innovate and to find solutions outside the box.
This balance cannot be obtained by brute force, in a top-down approach to system dynamics, it has to be found in a bottom-up approach that focuses on the system elements and their interactions.

Our paper contributes to this discussion in several ways.
We study a problem of practical relevance that can hardly be solved in a top-down approach: the collapse of an online social network (OSN) because the decision of some users to leave causes the drop-out
 of others at large scale.  
A real-world example is the collapse of the OSN \emph{Friendster} \citep{Garcia2013}. 
As long as users are free to stay or to leave, the \emph{emergence}, of such large failure cascades cannot be prevented by administrative ruling.
Applying global incentives for users to stay, on the other hand, usually implies high costs and questionable efficiency.

Therefore, in this paper, we propose a bottom-up approach to influence the OSN on the level of users, i.e., agents in our model.
They can be targeted in two ways: by influencing their interactions or by influencing their utility.
We have argued for the latter, because of the large volatility in the dynamics of the OSN.
Specifically, we propose to change the costs of particular agents such that the overall robustness of the OSN is increased.
As already mentioned in the Introduction, OSN should be seen as \emph{socio-technical systems}, and it is in fact the \emph{technical} component that in principle allows us to influence the costs of users much easier than it would be possible in the offline world. 

Improving robustness first requires us to define an appropriate measure of robustness suitable for real-world OSN.
Here we propose the \emph{average in-degree coreness}, which does not just reflect the degree of agents but quantifies how well they are \emph{integrated} in the OSN.
Next, we have to understand why robustness \emph{decreases} in the absence of network interventions.
Based on computer simulations and detailed discussions of agent benefits and costs, we show that it is the changing relation between the core and the periphery of the OSN, which eventually destabilizes the network.

Based on these insights, we have proposed two different scenarios for network interventions to improve robustness.
The first one targets peripheral agents and reduces their cost, to incentivize them to \emph{stay} in the OSN.
The second one targets only \emph{one} agent from a $k$-shell next to the core and increases its cost, to incentivize it to \emph{leave} the OSN.
Both scenarios have in common to increase the size of the periphery, but they reach this goal in different ways.
As we demonstrate by means of computer simulations, the first scenario is able to considerably \emph{delay} the breakdown of the OSN, while the second one is able to prevent this breakdown.
Dependent on the optimal choice of parameters, we could show that even the peripheral control improves the robustness of the OSN in a \emph{statistically significant} manner.
Still, we argue that the second scenario should be the preferred one because it requires  (i) to only control a single agent instead of many, and (ii) less investment because, instead of \emph{decreasing} the costs of many agents via compensations, here the cost is increased.

Our findings are interesting and, at first sight, also counter-intuitive because 
they challenge our understanding of how to improve the robustness of systems.
One could simply argue that the best way to increase robustness is to keep all parts of the system tightly together, to not lose anything.
This may apply to mechanical or technical systems.
But for socio-technical and socio-economic systems, we have to take into account their adaptivity and their ability to respond to changes in an unintended manner.  
Therefore, the first step for interventions is to understand the eigendynamics of these systems, i.e., their behavior in the absence of regulations or control.
To achieve this understanding in the case of complex systems, agent-based modeling is the most appropriate way.
Different from a complex network approach that focuses mainly on the link topology, agent-based modeling allows also capturing the internal dynamics of the system elements, i.e., the nodes or agents, in response to interactions. 
Only this advanced level of modeling enables us to propose interventions targeted at specific agents and to investigate how the system as a whole responds to these network interventions.

\small \setlength{\bibsep}{1pt}


\begin{thebibliography}{15}
\expandafter\ifx\csname natexlab\endcsname\relax\def\natexlab#1{#1}\fi
\expandafter\ifx\csname url\endcsname\relax
  \def\url#1{\texttt{#1}}\fi
\expandafter\ifx\csname urlprefix\endcsname\relax\def\urlprefix{URL }\fi
\expandafter\ifx\csname selectlanguage\endcsname\relax
  \def\selectlanguage#1{\relax}\fi

\bibitem[{Borgatti and Everett(2000)}]{borgatti2000models}
Borgatti, S.~P.; Everett, M.~G. (2000).
\newblock Models of core/periphery structures.
\newblock \emph{Social networks} \textbf{21(4)}, 375--395.

\bibitem[{Garcia \emph{et~al.}(2013)Garcia, Mavrodiev and
  Schweitzer}]{Garcia2013}
Garcia, D.; Mavrodiev, P.; Schweitzer, F. (2013).
\newblock Social resilience in online communities: The autopsy of Friendster.
\newblock In: \emph{1st ACM Conference in Online Social Networks (COSN'13)}.
  pp. 39--50.

\bibitem[{Jain and Krishna(1998)}]{Jain1998}
Jain, S.; Krishna, S. (1998).
\newblock {Emergence and Growth of Complex Networks in Adaptive Systems}.
\newblock \emph{Computer Physics Communications} \textbf{122}, 10.

\bibitem[{Jain and Krishna(2002)}]{Jain2002}
Jain, S.; Krishna, S. (2002).
\newblock {Crashes, Recoveries, and Core Shifts in a Model of Evolving
  Networks}.
\newblock \emph{Physical Review} \textbf{65(2)}, 26103--26104.

\bibitem[{Kairam \emph{et~al.}(2012)Kairam, Wang and Leskovec}]{Kairam2012}
Kairam, S.~R.; Wang, D.~J.; Leskovec, J. (2012).
\newblock {The life and death of online groups}.
\newblock In: \emph{Proceedings of the fifth ACM international conference on
  Web search and data mining - WSDM '12}. New York, New York, USA: ACM Press,
  p. 673.

\bibitem[{Liu \emph{et~al.}(2011)Liu, Slotine and Barab{\'a}si}]{Liu2011}
Liu, Y.-Y.; Slotine, J.-J.; Barab{\'a}si, A.-L. (2011).
\newblock Controllability of complex networks.
\newblock \emph{Nature} \textbf{473(7346)}, 167.

\bibitem[{{\L}uczak(1991)}]{Luczak1991}
{\L}uczak, T. (1991).
\newblock {Size and connectivity of the k-core of a random graph}.
\newblock \emph{Discrete Mathematics} \textbf{91(1)}, 61--68.

\bibitem[{Schweitzer(1997)}]{fs-ed-97}
Schweitzer, F. (ed.) (1997).
\newblock \emph{Self-Organization of Complex Structures: From Individual to
  Collective Dynamics. Part 1: Evolution of Complexity and Evolutionary
  Optimization, Part 2: Biological and Ecological Dynamics, Socio-Economic
  Processes, Urban Structure Formation and Traffic Dynamics}.
\newblock London: Gordon and Breach.

\bibitem[{Schweitzer(2019)}]{Schweitzer:2019vp}
Schweitzer, F. (2019).
\newblock The Bigger Picture: Complexity Meets Systems Design.
\newblock In: G.~Folkers; M.~Schmid (eds.), \emph{Design. Tales of Science and
  Innovation}, Chronos Verlag, Zurich. pp. 77 -- 86.

\bibitem[{Schweitzer \emph{et~al.}(2019{\natexlab{a}})Schweitzer, Casiraghi and
  Perony}]{Perony2019}
Schweitzer, F.; Casiraghi, G.; Perony, N. (2019{\natexlab{a}}).
\newblock Modeling the emergence of hierarchies from dominance interactions.
\newblock \emph{Bulletin of Mathematical Biology} , (submitted).

\bibitem[{Schweitzer \emph{et~al.}(2019{\natexlab{b}})Schweitzer, Mavrodiev,
  Seufert and Garcia}]{mavrodiev2019}
Schweitzer, F.; Mavrodiev, P.; Seufert, A.~M.; Garcia, D. (2019{\natexlab{b}}).
\newblock Modeling User Reputation in Online Social Networks: The Role of
  Costs, Benefits, and Reciprocity.
\newblock \emph{Computational and Mathematical Organization Theory} ,
  (submitted).

\bibitem[{Seidman(1983)}]{Seidman1983a}
Seidman, S.~B. (1983).
\newblock {Network structure and minimum degree}.
\newblock \emph{Social Networks} \textbf{5(3)}, 269--287.

\bibitem[{Wasserman and Faust(1994)}]{Wasserman1994}
Wasserman, S.; Faust, K. (1994).
\newblock \emph{Social network analysis: Methods and applications}.
\newblock Cambridge university press.

\bibitem[{Zhang \emph{et~al.}(2016)Zhang, Garas and Schweitzer}]{Zhang2016}
Zhang, Y.; Garas, A.; Schweitzer, F. (2016).
\newblock Value of peripheral nodes in controlling multilayer scale-free
  networks.
\newblock \emph{Phys. Rev. E} \textbf{93}, 012309.

\bibitem[{Zhang \emph{et~al.}(2019)Zhang, Garas and
  Schweitzer}]{zhang2019control}
Zhang, Y.; Garas, A.; Schweitzer, F. (2019).
\newblock Control contribution identifies top driver nodes in complex networks.
\newblock \emph{arXiv preprint arXiv:1906.04663} .

\end{thebibliography}
\end{document}